\documentclass[aps,prb,twocolumn,superscriptaddress,showpacs]{revtex4-1}
\usepackage[colorlinks, linkcolor=blue, anchorcolor=blue, citecolor=blue]{hyperref}
\usepackage{amsmath}
\usepackage{graphicx}
\usepackage{braket}
\usepackage{color}

\begin{document}

\title{Interaction-induced quantum anomalous Hall phase in (111) bilayer of LaCoO$_3$ }

\author{Yilin Wang,$^{1}$ Zhijun Wang,$^{1}$ Zhong Fang,$^{1,2}$ and Xi Dai}

\affiliation{ Beijing National Laboratory for Condensed Matter Physics, 
              and Institute of Physics, 
              Chinese Academy of Sciences, 
              Beijing 100190, 
              China }

\affiliation{Collaborative Innovation Center of Quantum Matter, Beijing
100190, China}

\date{\today}

\begin{abstract}
In the present paper, the Gutzwiller density functional theory (LDA+G) has been applied to study 
the bilayer system of LaCoO$_3$ grown along the $(111)$ direction on SrTiO$_3$.
The LDA calculations show that there are two nearly flat bands located at the top and bottom of $e_{g}$ 
bands of Co atoms with the Fermi level crossing the lower one. 
After including both the spin-orbit coupling and the Coulomb interaction in the LDA+G method, 
we find that the interplay between spin-orbit coupling and Coulomb interaction stabilizes a very robust ferromagnetic
insulator phase with non-zero Chern number indicating the possibility to realize quantum 
anomalous Hall effect in this system.
\end{abstract}


\maketitle

%% introduction
% intro 1: searching topological phase in oxides, and the advantages
\section{introduction}\label{sec:intro}
Searching for topologically non-trivial phases~\cite{hasan:2010,qi:2011} in realistic material systems is one of the
fast developing research fields in condensed matter physics. 
Recently, transition metal oxides (TMOs) have been proposed~\cite{xiao:2011} as a potential platform 
for topological materials due to many of its advantages compared to the previously found topological compounds, 
such as HgTe~\cite{bernevig:2006,konig:2007,dai:2008} and Bi$_2$Se$_3$ family compounds~\cite{zhang:2009,xia:2009,chen:2009}. 
First of all, oxides are chemically much more stable when exposing in the air, which makes it more attractive 
for potential applications. Secondly, the relatively strong Coulomb interaction among the electrons in $3d$ orbitals  
generates fruitful many-body physics in TMOs and provides a tantalizing field for the search of topological materials.
Finally, the rapid development of the techniques for the heterostructure growth of TMOs~\cite{izumi:2001,ohtomo:2002,ohtomo:2004}  
paves a completely new path to realize topological phases in condensed matter systems by material design. 

% intro 2: (111) bilayer, honeycomb lattice, LaAuO3, LaNiO3
D. Xiao \emph{et al.}~\cite{xiao:2011} first pointed out that the (111) bilayer heterostructure of perovskite TMOs
can be viewed as a buckled honeycomb lattice and has ``natively" inverted band structure, which is very similar 
to the situation in graphene~\cite{castro:2009}. Based on the tight-binding (TB) and the first-principles 
calculations of (111) bilayer heterostructure of TMOs, they found that there are two nearly flat bands at the 
top and bottom of the $e_{g}$ bands of TM ions, together with another two bands with nearly linear dispersion 
forming a similar Dirac point at $K$ point. The appearance of both flat bands and Dirac points in this system 
can be ascribed to the special geometry effects of the honeycomb lattice, based on which several exotic 
topological states can be designed. The first one is to open a semiconductor gap at the Dirac point by  
spin-orbit coupling (SOC) leading to quantum spin Hall effect. Because the sizable strength of SOC is required 
in this proposal, the typical realistic system proposed in Ref.~\onlinecite{xiao:2011} is LaAlO$_3$/LaAuO$_3$/LaAlO$_3$ 
heterostructure, which is still very difficult to grow experimentally so far. The second proposal in this field 
is to realize quantum anomalous Hall effect (QAHE) in the (111) bilayer of $3d$ TMOs~\cite{yang:2011,andreas:2011,andreas:2012,andreas:2013},
which is experimentally much more feasible than the $4d$ or $5d$ TMOs. Although the SOC strength is about one order 
smaller than that of $4d$ or $5d$ TMOs, the strong Coulomb interaction can help to stabilize the topologically 
non-trivial phases. Recently, the similar heterostructure has been made, but unfortunately the ground state is found to be
anti-ferromagnetic (AFM)~\cite{middey:2014}. 

% intro 3: our propose, (111) bilayer of LaCoO3 grown on SrTiO3
In this paper, we propose another bilayer heterostructure of $3d$ TMO, LaCoO$_3$, grown along the (111) direction 
on SrTiO$_3$, as a promising candidate to realize QAHE, which has the following advantages: 
(1) comparing with $4d$ and $5d$ TMOs, LaCoO$_3$ is more accessible and its thin film is not difficult to synthesize,
(2) the lower flat band in the heterostructure leads to a sharp peak of density of states around the 
Fermi level which induces a strong Stoner instability towards the ferromagnetic (FM) insulator phase,
(3) although the bare SOC strength is quite small, it can be largely enhanced by strong Coulomb interaction 
among $3d$ electrons,
(4) with both large enough exchange splitting and modified SOC induced by strong correlation effects, the system 
falls into a semiconductor state under FM order with a non-zero total Chern number.
	
%% method
\section{method}\label{sec:method}
LaCoO$_3$ is a typical material with its band structure strongly modified by correlation effects. 
In this work, the Gutzwiller density functional theory (LDA+G)~\cite{bunemann:1998,deng:2009, nicola:2012, lu:2013} 
is used to calculate the ground states and quasi-particle band structures of its bulk phase and heterostructure. 
The LDA part of calculations have been done by the Vienna Ab-initio Simulation Package (VASP)~\cite{kresse:1996} 
with projector augmented-wave (PAW) pseudopotential~\cite{blochl:1994,kresse:1999} and Perdew-Burke-Ernzerhof parametrization 
of the generalized gradient approximation (GGA-PBE) exchange-correlation functionals~\cite{perdew:1996}. 
The energy cutoff of the plane-wave basis is set to be 400 eV, and a $\Gamma$-centered $11 \times 11 \times 11$ $K$-point grid
for the bulk and $8 \times 8 \times 1$ for the heterostructure has been chosen, respectively.

In the Gutzwiller part, we solve a Hamiltonian in the Wannier representation, which reads
\begin{eqnarray}
\label{eqn:ham}
 H &=& \sum_{ij,\alpha\beta\sigma}(t_{ij,\alpha\beta\sigma}^{dd} d_{\alpha\sigma}^{\dagger}d_{\beta\sigma}
      +t_{ij,\alpha\beta\sigma}^{pp} p_{\alpha\sigma}^{\dagger}p_{\beta\sigma} \nonumber \\
   & &+t_{ij,\alpha\beta\sigma}^{dp} d_{\alpha\sigma}^{\dagger}p_{\beta\sigma} 
      + t_{ij,\alpha\beta\sigma}^{pd} p_{\alpha\sigma}^{\dagger}d_{\beta\sigma} ) \nonumber \\ 
   & &+ \frac{1}{2}\sum_{i,\alpha\beta\gamma\delta\sigma\sigma^{\prime}} U_{\alpha\sigma,\beta\sigma^{\prime},
        \gamma\sigma,\delta\sigma^{\prime}}d_{\alpha\sigma}^{\dagger}d_{\beta\sigma^{\prime}}^{\dagger}
         d_{\delta\sigma^{\prime}}d_{\gamma\sigma} \nonumber \\
   & &-\sum_{i,\alpha\sigma} \bar{U}(n_{d}-\frac{1}{2})d_{\alpha\sigma}^{\dagger}d_{\alpha\sigma},
\end{eqnarray}
where $i,j$ is the site index, $\alpha,\beta,\gamma,\delta$ is the Wannier orbital index, $\sigma,\sigma^{\prime}$ 
is the spin index.

The first two lines of Eqn.~\ref{eqn:ham} describe a $d$-$p$ TB Hamiltonian consists of Co $3d$ orbitals and O $2p$ orbitals, 
which are constructed from the non-SOC LDA calculation by the maximally localized Wannier functions (MLWF) method~\cite{marzari:2012} 
implemented in the WANNIER90~\cite{mostofi:2008} package.

The third line of Eqn.~\ref{eqn:ham} describes the local atomic Coulomb interaction for Co $3d$ orbitals. We assume the spherical symmetry for
Coulomb interaction in LaCoO$_3$ solid and use a full interaction tensor 
$U_{\alpha\sigma,\beta\sigma^{\prime},\gamma\sigma,\delta\sigma^{\prime}}$ for the entire $d$-shell~\cite{sugano:1970,georges:2013}.
We first write down the $U$-tensor in the complex spherical harmonics basis $\phi_{m}=R_{3d}(r)Y_{2}^{m}$. In this basis, 
the $U$-tensor is
\begin{equation}
\label{eqn:utensor}
U_{m_{1}\sigma,m_{2}\sigma^{\prime},m_{3}\sigma,m_{4}\sigma^{\prime}}=\delta_{m_1+m_2,m_3+m_4}\sum_{k}c_{k}^{m_1,m_3}c_{k}^{m_4,m_2}F^{k},
\end{equation}
where, $k=0,2,4$ for $d$-shell, $c_{k}^{m_1,m_3}$ is the Gaunt coefficient which has been exactly calculated and 
tabulated in Table 1.2 in Ref.~\onlinecite{sugano:1970}, and $F^{0},F^{2},F^{4}$ are the three Slater integrals which
are unknown. Thus, the full $U$-tensor is parameterized by $F^{0},F^{2},F^{4}$, however, we take $F^{4}/F^{2}=0.625$ as 
an approximation with good accuracy for $d$-shell~\cite{groot:1990}. Then we transform it to the Wannier basis (the cubic
spherical harmonics) by using 
the transformations: $d_{xy}=-\frac{i}{\sqrt{2}}(\phi_{2}-\phi_{-2}), d_{xz}=-\frac{1}{\sqrt{2}}(\phi_{1}-\phi_{-1}),
d_{yz}=\frac{i}{\sqrt{2}}(\phi_{1}+\phi_{-1}), d_{x^2-y^2}=\frac{1}{\sqrt{2}}(\phi_{2}+\phi_{-2}), d_{3z^2-r^2}=\phi_{0}$. 
We don't follow the traditional definition of Coulomb interaction as $U_d=F^0$ and Hund's rule coupling as $J_H=\frac{1}{14}(F^2+F^4)$,
instead, we define the Kanamori type $U$ and $J$ in the Wannier basis~\cite{georges:2013}, where the intra-orbital Coulomb interaction is $U=F^{0}+\frac{4}{49}F^{2}+\frac{4}{49}F^{4}$,
and the Hund's rule coupling is orbital-dependent (anisotropic): 
$J(d_{xy},d_{xz})=J(d_{xy},d_{yz})=J(d_{xz},d_{yz})=J(d_{xz},d_{x^2-y^2})=J(d_{yz},d_{x^2-y^2})=\frac{3}{49}F^2+\frac{20}{441}F^4$,
$J(d_{xy},d_{3z^2-r^2})=J(d_{x^2-y^2},d_{3z^2-r^2})=\frac{4}{49}F^2+\frac{15}{441}F^4$,
$J(d_{xz},d_{3z^2-r^2})=J(d_{yz},d_{3z^2-r^2})=\frac{1}{49}F^2+\frac{30}{441}F^4$, 
$J(d_{xy},d_{x^2-y^2})=\frac{35}{441}F^4$. 
We take all the terms of Hund's rule coupling into account in our calculations and define an average value of them 
as $J=\frac{5}{98}(F^2+F^4)$ for the convenience of discussing our results. 
Thus, given parameters $F^{0},F^{2}$ or $U,J$, we can construct the full interaction $U$-tensor.

The fourth line of Eqn.~\ref{eqn:ham} is the double-counting term~\cite{anisimov:2010} used to substract the correlation effect which has been considered 
in the LDA calculations. {$\bar{U}$ is the average Coulomb interaction, which is defined as
\begin{equation}
\bar{U}=\frac{ \sum_{a}U + \sum_{a<b}(U-2J_{a,b}) + \sum_{a<b}(U-3J_{a,b}) }{M(2M-1)},
\end{equation} 
where, $1\leq a,b \leq M$, $M$ is the total number of orbitals (not including spin), $U$ is the intra-orbital Coulomb interaction, and
$J_{a,b}$ is the Hund's rule coupling between orbitals $a$ and $b$. 
$n_{d}$ is the total occupancy of $3d$ orbitals for one Co site from the LDA calculations.

This Hamiltonian will be treated by the rotationally invariant Gutzwiller variational method introduced in detail 
in Ref.~\onlinecite{bunemann:1998,deng:2009, nicola:2012, lu:2013}. The Gutzwiller variational wave function $\Ket{G}$ is 
constructed by applying a projector operator $P$ to the noninteracting wave function $\Ket{0}$ derived from
the LDA calculation,  
\begin{equation}
\Ket{G}=P\Ket{0}.
\end{equation}
The projector operator is chosen as
\begin{equation}
P=\prod_{i}(\sum_{\alpha}\lambda_{\alpha}^{i}\Ket{\Gamma_{\alpha}^{i}}\Bra{\Gamma_{\alpha}^{i}}),
\end{equation}
where $i$ is the site index, $\Ket{\Gamma}$ is the atomic eigenstates and $\lambda_{\alpha}$ are 
the Gutzwiller variational parameters, which can be determined by minimizing 
the total energy of the ground state.

\section{results and discussion}\label{res}

%% bulk of LaCoO3
% figure: the lattice structure and electronic structure of bulk LaCoO3
\begin{figure}
\includegraphics[width=0.45\textwidth]{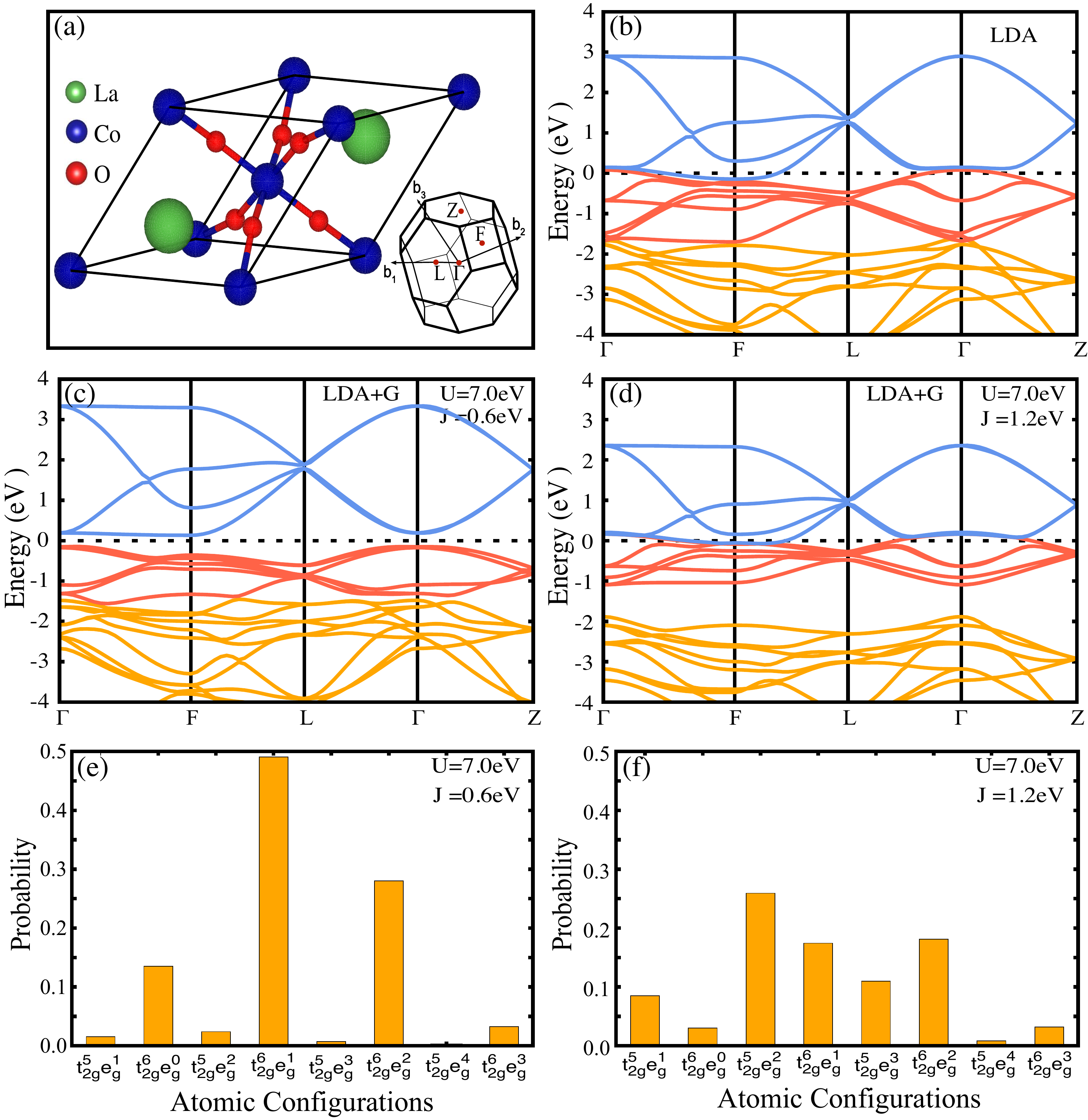}
\caption{(Color online). (a) The crystal structure of bulk LaCoO$_3$ with $R\bar{3}c$ (No. 167) space group and the 
         corresponding Brillouin zone. (b) is the band structures obtained from the LDA calculation. (c) and (d) are the band
         structure obtained from the LDA+G calculations with Coulomb interaction $U=7.0$ eV, Hund's rule coupling
         $J=0.6$ eV and $J=1.2$ eV, respectively. (e) and (f) are the corresponding probabilities of 
         the atomic configurations $\Ket{I}$
         in the Gutzwiller wave function $\Ket{G}$, $P_{I}=\Braket{G|I}\Braket{I|G}$ indicating the LS state for $J=0.6$ eV 
         and HS state for $J=1.2$ eV.} 
\label{fig:bulk}
\end{figure}

% figure: the partial density of states of bulk LaCoO3
\begin{figure}
\includegraphics[width=0.45\textwidth]{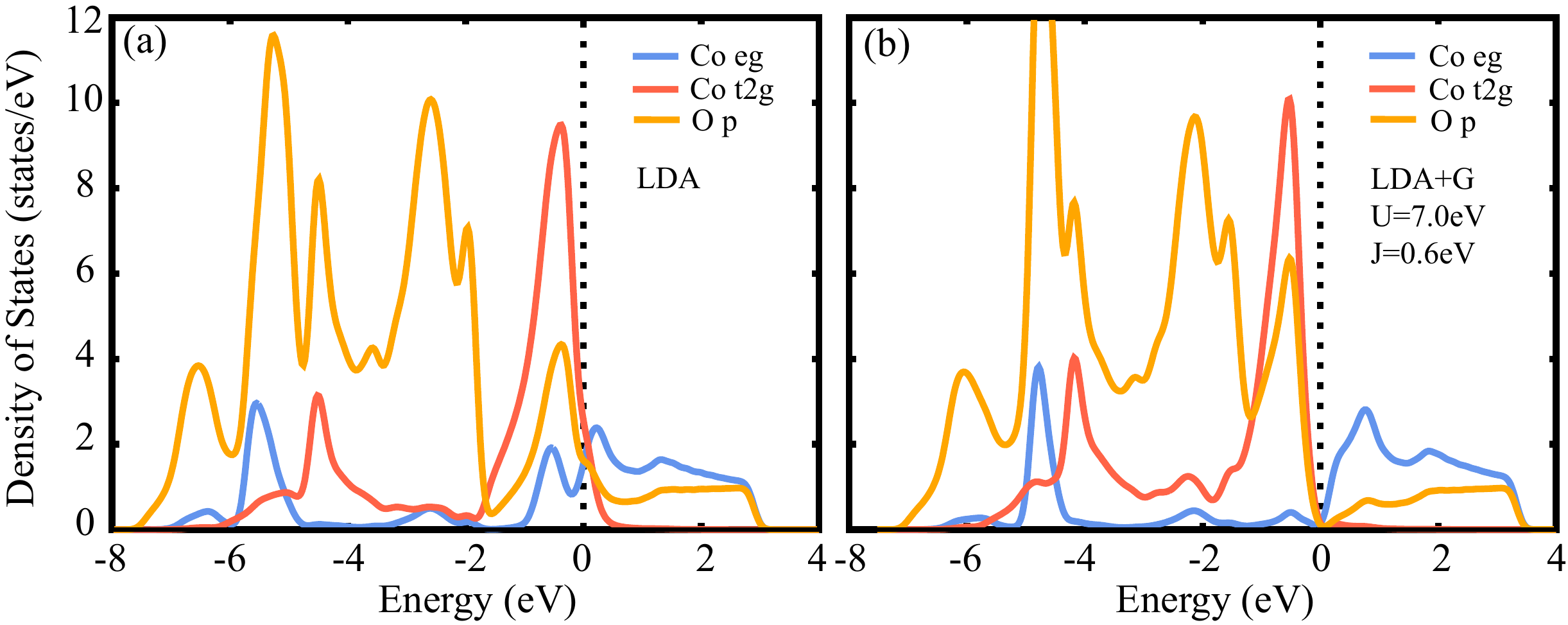}
\caption{(Color online). The partial density of states for total Co $e_{g}$, $t_{2g}$ states and
         total O $2p$ states in one primitive cell of bulk LaCoO$_3$, (a) for LDA calculation and 
         (b) for LDA+G calculation with $U=7.0$ eV, $J=0.6$ eV.}
\label{fig:dos}
\end{figure}

The bulk material of LaCoO$_3$ has very complicated electronic and spin state transitions~\cite{abbate:1994,
korotin:1996,yamaguchi:1997,asai:1998,yamaguchi:1996,tokura:1998,imada:1998,radaelli:2002,zobel:2002,
knizek:2005,haverkort:2006,klie:2007,pandey:2008,craco:2008,hsu:2009,eder:2010,kune:2011,krapek:2012,zhang:2012}.
At low enough temperature ($T<50$ K), it is a semiconductor with low spin (LS) state. With the increasing of temperature,
it undergoes a spin state transition to intermediate spin (IS) state around $T=100$ K. When temperature $T>500$ K, 
another transition from IS semiconductor to high spin (HS) metal will occur. 
However, we just focus on the zero-temperature ground state in our calculations.
As shown in Fig.~\ref{fig:bulk}(a), the bulk 
LaCoO$_3$ has a distorted perovskite structure with $R\bar{3}c$ (No. 167) space group~\cite{radaelli:2002,krapek:2012}, 
which contains two equivalent Co atoms in each unit cell. In our calculations, we take the lattice parameters for 
temperature $T=5$ K from Ref.~\onlinecite{radaelli:2002}.  

The LDA band structure is shown in Fig.~\ref{fig:bulk}(b). 
The $e_{g}$ (blue) and $t_{2g}$ (red) bands of Co atoms overlap and give a metallic ground state contradicting 
with the experimental results~\cite{yamaguchi:1996,tokura:1998,imada:1998} which show semiconductor behavior. 
After considering the Coulomb interaction and the Hund's rule coupling in our LDA+G calculations, 
we get a semiconductor ground state when $U=7.0$ eV and $J=0.6$ eV, 
as shown in Fig.~\ref{fig:bulk}(c), which is in good agreement with both the experimental result~\cite{yamaguchi:1996,tokura:1998,imada:1998}
and the numerical result obtained by LDA+DMFT~\cite{krapek:2012}. 
Comparing with the LDA band structure in Fig.~\ref{fig:bulk}(b),
the Gutzwiller method modifies the bands in two ways: (1) renormalizes the effective crystal field splitting between $t_{2g}$ and $e_g$ orbitals; and 
(2) renormalizes the bandwidth of $t_{2g}$ and $e_{g}$ bands to be much narrower, as a result, opens a gap between them. 
We also obtain the partial density of states from the LDA and LDA+G calculations, which are plotted in Fig.~\ref{fig:dos}(a,b). 
As we can see, the Co $3d$ bands strongly hybridize with O $2p$ bands, as a result, the calculated total occupation number of Co $3d$ orbitals,
which is about 7.2, is larger than the nominal one 6. From the quasi-particle band structure obtained by LDA+G, we can find that
the band gap is between $e_g$ and $t_{2g}$ and it is
a typical semiconductor with its band width renormalized by interaction about $80\%$.
The corresponding probability of the atomic configurations $\Ket{I}$ in the Gutzwiller ground 
state $\Ket{G}$ can be calculated using the Gutzwiller wave function
as $P_{I}=\Braket{G|I}\Braket{I|G}$, which are plotted in Fig.~\ref{fig:bulk}(e). 
There are mainly three configurations
$t_{2g}^{6}e_{g}^{0},t_{2g}^{6}e_{g}^{1},t_{2g}^{6}e_{g}^{2}$ in the ground state indicating a LS state.   
Note that we call the spin states with fully filled $t_{2g}$ orbitals ($t_{2g}^{6}$) as LS states, while those with some  
holes in the $t_{2g}$ orbitals as HS states.
To check how the Hund's rule coupling affects the spin states and the electronic structure, we increase
it to be $J=1.2$ eV, and we get a metallic electronic structure with HS state,  
as shown in Figs.~\ref{fig:bulk}(d, f).
Locking between the metal-semi-conductor transition with the spin state
transition can be explained by the competition between the cubic crystal splitting and Hund's rule coupling. As
a consequence, the increasing of Hund's rule coupling  will strongly suppress the effective crystal splitting 
between $t_{2g}$ and $e_g$ orbitals leading to the vanish of the semiconductor gap between them.
We want to emphasize that this HS state is still a zero-temperature ground state, which is different with 
the temperature induced IS and HS states in LaCoO$_3$.

%% heterostructure of LaCoO3

% figure: the lattice structure and LDA+SOC band structure of heterostructure LaCoO3
\begin{figure}
\includegraphics[width=0.54\textwidth]{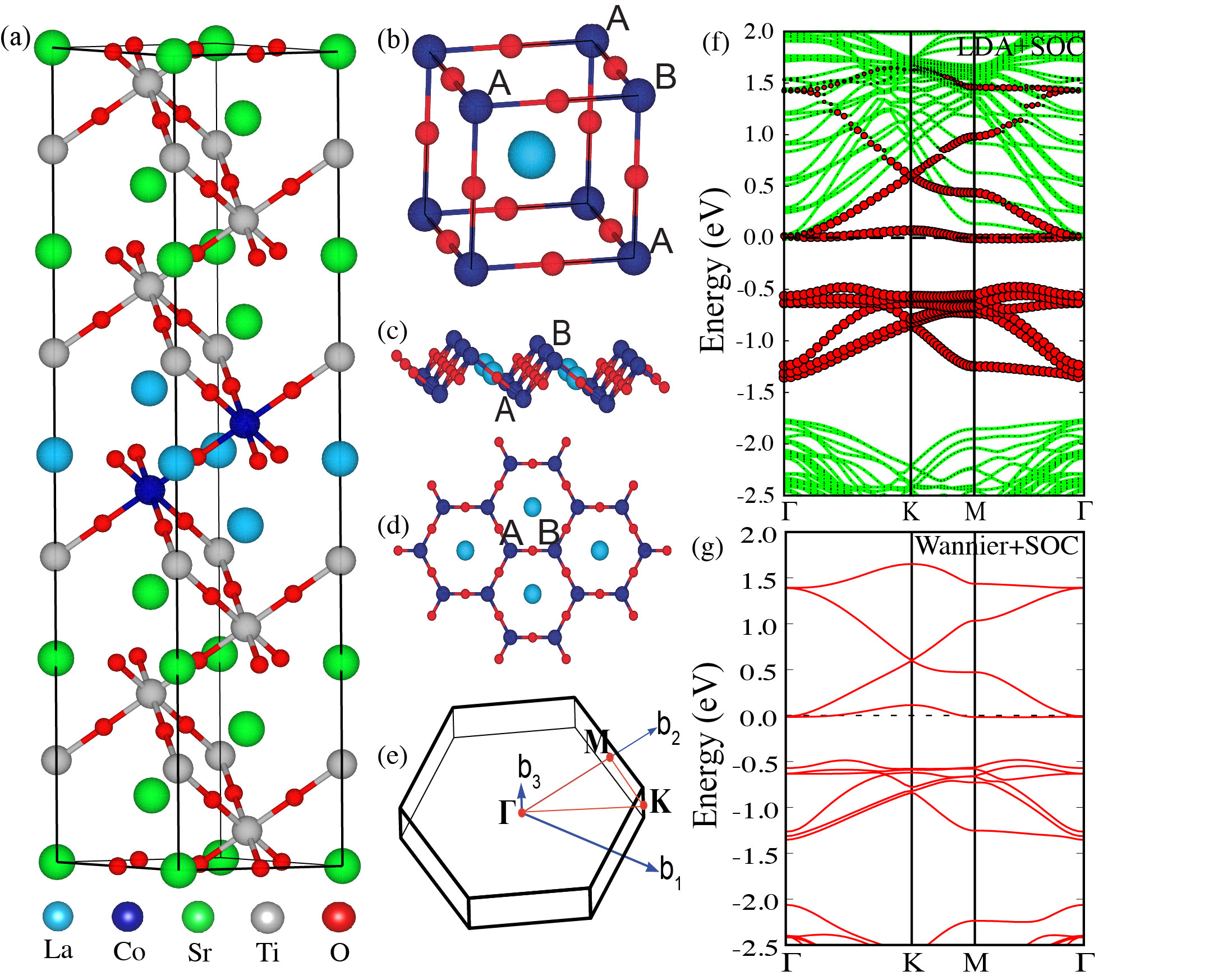}
\caption{(Color online). (a) The heterostructure of LaCoO$_3$, two layers of Co and three layers of LaO$_3$ are
         grown along the (111) direction on SrTiO$_3$. (b, c, d) illustrate the formation of a buckled honeycomb lattice
         by two layers of TM ions along the (111) direction of an ideal perovskite lattice. (e) is the Brillouin 
         zone of the honeycomb lattice. (f) is the fat bands derived from the LDA+SOC calculation and (g) is the 
         band structure obtained by the Wannier TB Hamiltonian.}
\label{fig:hetero}
\end{figure}

% figure: the charge transfer
\begin{figure}
\includegraphics[width=0.26\textwidth]{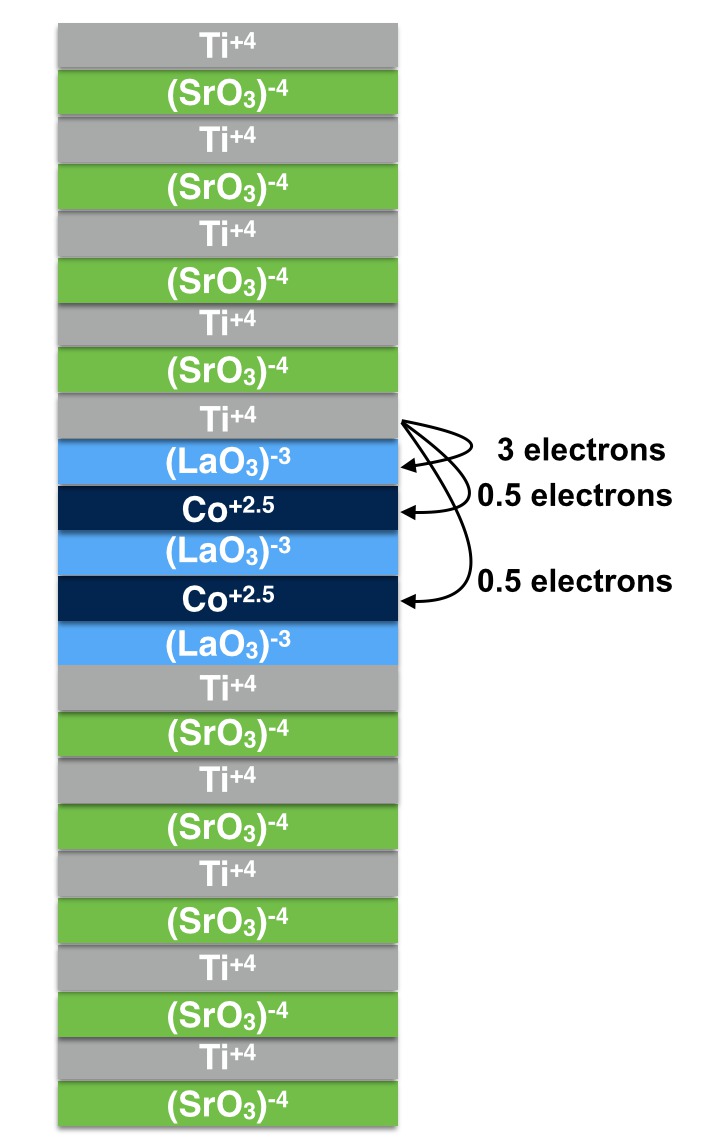}
\caption{(Color online). Illustration of the charge transfer from Ti to Co in the heterostructure of LaCoO$_3$}
\label{fig:charge}
\end{figure}

% figure: phase diagram 
\begin{figure}
\includegraphics[width=0.48\textwidth]{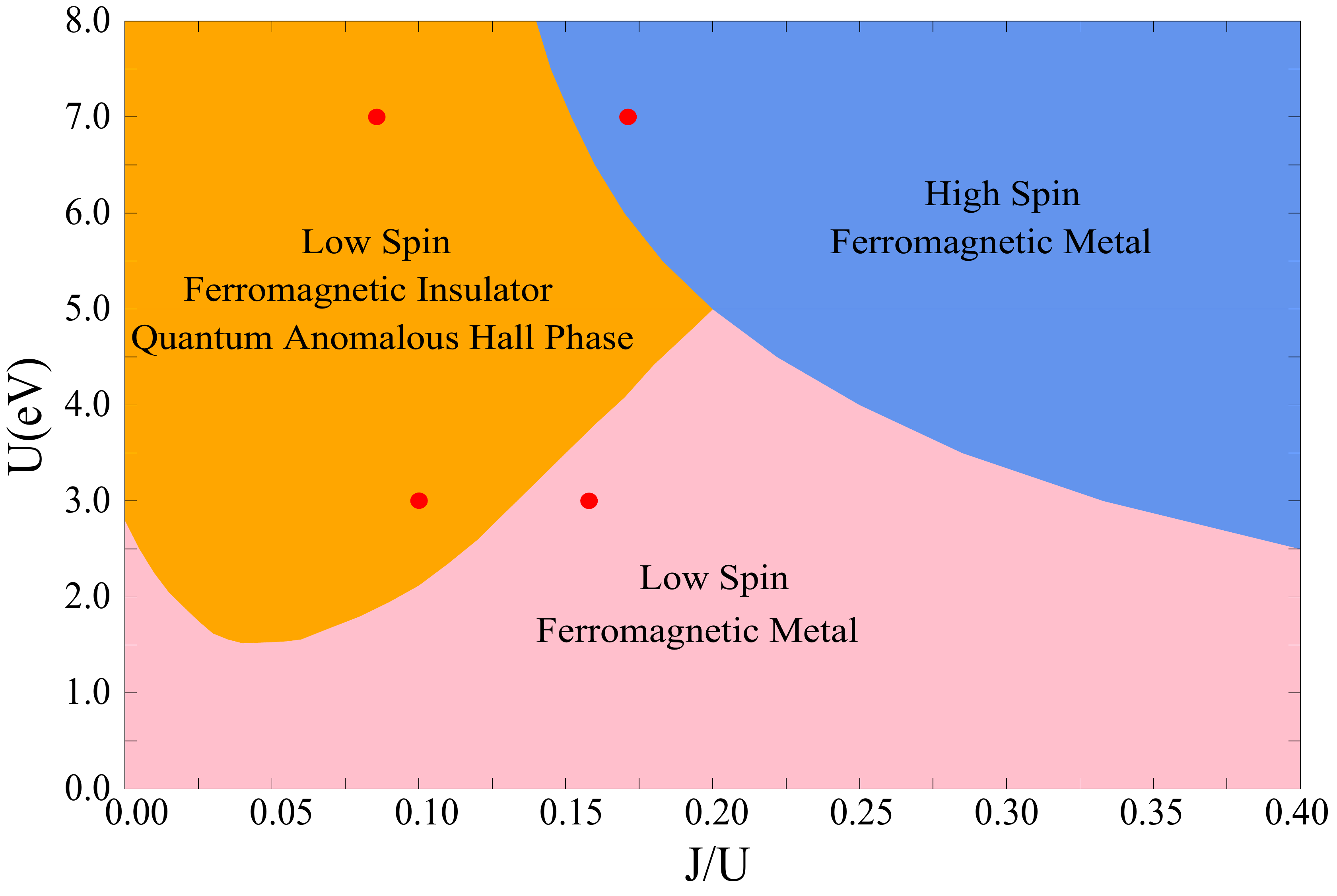}
\caption{(Color online). The phase diagram calculated by the LDA+G method in the plane of Coulomb interaction $U$ and 
          Hund's rule coupling $J$. There are three regions: (pink) low spin ferromagnetic metal (LS-FM-M), (orange) 
          low spin ferromagnetic insulator (LS-FM-I), (blue) high spin ferromagnetic metal (HS-FM-M).}
\label{fig:phase}
\end{figure}

% figure: the polarization
\begin{figure}
\includegraphics[width=0.48\textwidth]{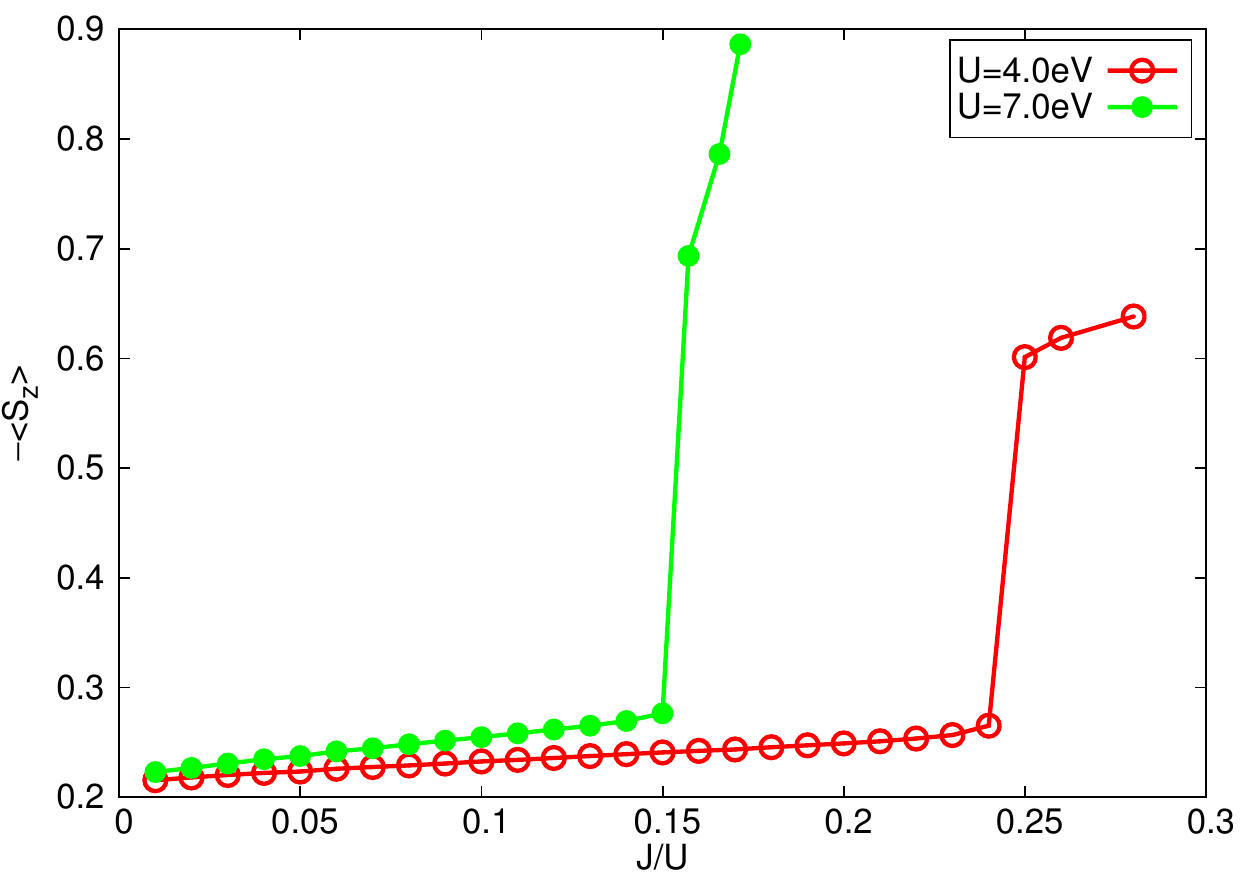}
\caption{(Color online). The magnetization of the ground state as the function of $J/U$ for two different Hubbard $U$ value.}
\label{fig:sz}
\end{figure}

% figure: the band structure and multiplet weight of the heterostructure 
\begin{figure*}
\includegraphics[width=0.90\textwidth]{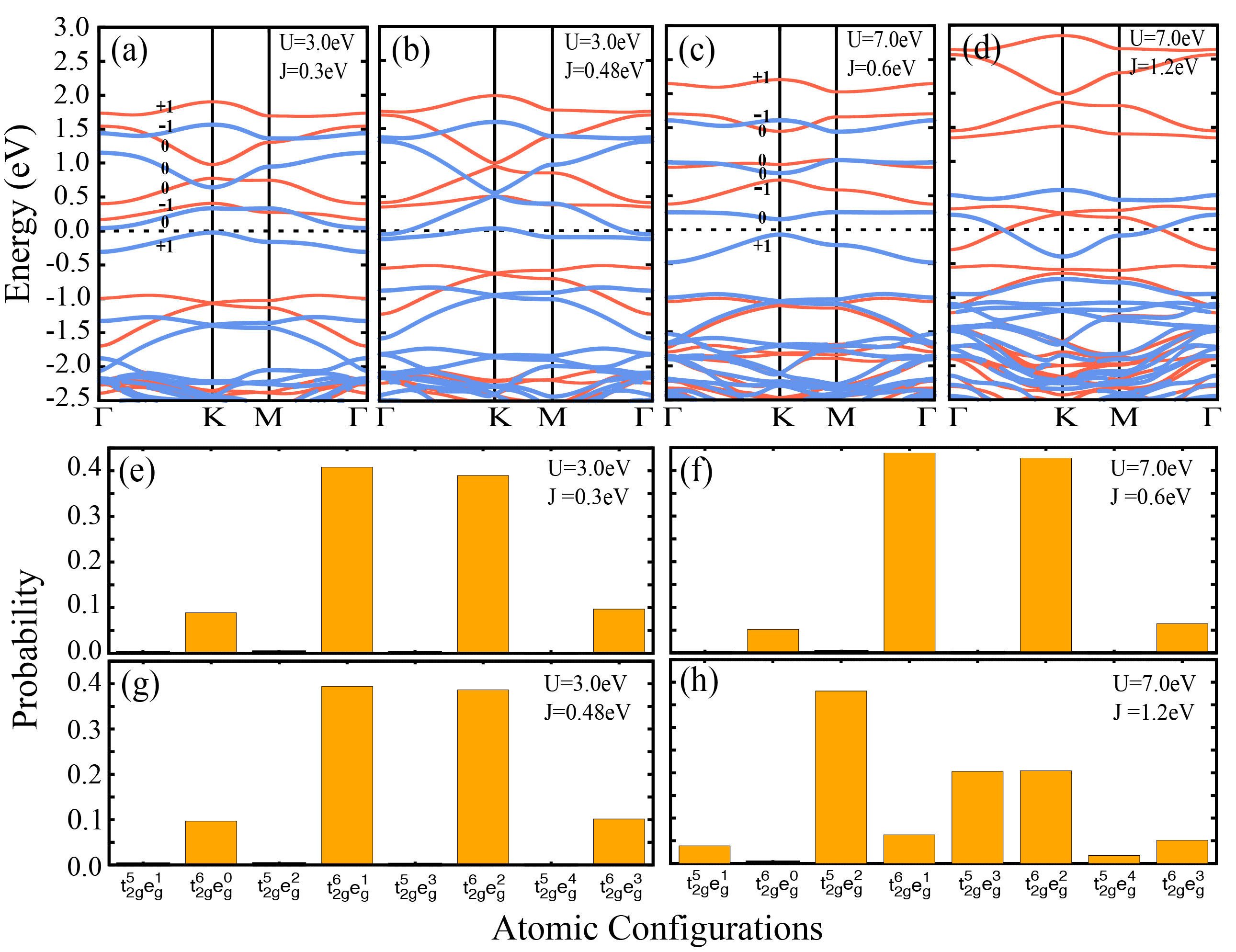}
\caption{(Color online). Four points are chosen from the phase diagram to plot their band structures 
         and corresponding probability of the atomic configurations. (a, b, c, d) The Gutzwiller quasi-particle 
         band structures, blue and red lines indicate the majority and minority bands respectively, the numbers
         in (a, c) are the Chern numbers for separate band. (e, f, g, h) The corresponding probability of the 
         atomic configurations $\Ket{I}$ in the Gutzwiller wave function $\Ket{G}$ indicating the spin states. }
\label{fig:band}
\end{figure*}

We now turn to the heterostructure of LaCoO$_3$. The heterostructure of LaCoO$_3$ proposed in this paper is shown 
in Fig.~\ref{fig:hetero}(a), which contains two layers of Co and three layers of LaO$_3$ along the (111) direction 
on SrTiO$_3$. 
For one unit cell, the chemical formula is Ti$_{10}^{+4}$Co$_{2}^{+2.5}$(LaO$_3$)$_{3}^{-3}$(SrO$_3$)$_{9}^{-4}$.
The heterostructure belongs to space group $P\bar{3}M1$ (No. 164) with an inversion center located at 
O (0.5, 0.5, 0.5) site connecting two different layers of Co atoms, which form a buckled honeycomb lattice, as illustrated in 
Figs.~\ref{fig:hetero}(b, c, d). The lattice parameter of SrTiO$_3$ is fixed to be $3.95$ \AA~\cite{xiao:2011} and 
the internal atomic positions are optimized by LDA calculation using VASP. The LDA band structure with the optimized 
structure is then obtained and shown in Fig.~\ref{fig:hetero}(f), from which we can find two nearly flat bands with 
band width being around $0.06$ eV at the top and the bottom of the Co $e_{g}$ bands. The most important difference 
between the electronic structure of the bulk material and heterostructure of LaCoO$_3$ is that the formal charge 
of Co $3d$ orbitals increases from $6$ to $6.5$ in the heterostructure. 
This is due to the fact that the (111) interface 
between SrTiO$_3$ and LaCoO$_3$ is polarized leading to one electron transfer from Ti to Co. As illustrated in Fig.~\ref{fig:charge},
there is charge mismatch between one layer of Ti$^{+4}$ and one layer of (LaO$_3$)$^{-3}$, Ti will lose 4 electrons, but 
LaO$_3$ can only accept 3 electrons, so there will be one electron left and will be transfered to the two layers of Co due 
to the conservation of total charge. We want to emphasize that this charge transfer is due to the electric polarization 
instead of the hybridization of bands.
As a consequence,the bottom flat band crosses the Fermi level and is nearly half-filled. The TB model used for the further study is then constructed 
in the following way. First, based on the non-SOC LDA calculations we construct a TB model containing 20 $d$-orbitals 
(including spin degree of freedom) from two Co atoms and 54 $p$-orbitals from the Oxygen atoms in the three nearest layers. 
Then an atomic SOC Hamiltonian is added to the Hamiltonian in Eqn.~\ref{eqn:ham}, which reads,
\begin{equation}
H_{\text{SOC}} = \lambda \sum_{i} \vec{\textbf{l}}_{i}\cdot\vec{\textbf{s}}_i,
\end{equation}
where the strength $\lambda=50$ meV is determined by fitting the LDA+SOC results.

The band structure without SOC show a clear quadratic band touching at the $\Gamma$ point, which is 
unstable against infinitesimally small Coulomb interaction if the chemical potential lies exactly at the touching point~\cite{sunkai:2009}.
With SOC, the four-fold degenerate bands at $\Gamma$ point split into two doubly degenerate bands by the second 
order effects of SOC. The splitting is around 7 meV in LaCoO$_3$, which can be hardly seen from Figs.~\ref{fig:hetero}(f, g). 
If the flat band is fully filled, the Berry phase structure generated by SOC around $\Gamma$ point makes it a 2D 
topological insulator with non-trivial $Z_2$ index. As mentioned above, in reality the flat band is only half-filled 
and the system will keep metallic in paramagnetic phase. While as we have already seen in the bulk calculation, 
the correct semiconductor like electronic states can only be obtained when the strong Coulomb interaction among the 
$3d$ electrons has been considered in a proper way. For the bilayer LaCoO$_3$ system, the Coulomb interaction has two 
important effects. First it greatly enhances the SOC, which stabilizes the topological phase. Secondly, it generate 
strong Stoner instability in the flat band and makes the ground state to be FM. When the FM exchange 
coupling overcomes the band width of the flat band, the flat band will be completely polarized and a FM insulator phase 
is formed with the gap opening at the $\Gamma$ point by correlation enhanced effective SOC. The non-magnetic
band structure is already Z$_2$ non-trivial, the spin up and down subsystems can be viewed as two Chern insulators
with opposite Chern number -1 and 1. Therefore when it is completely polarized,
it naturally leads to Chern insulator phase with Chern number 1. 

We then apply the LDA+G method to carefully consider the Coulomb interaction. A phase diagram with
both Coulomb interaction $U$ and Hund's rule coupling $J$ has been obtained as shown in Fig.~\ref{fig:phase}.
The phase diagram is obtained by searching for the ground state by LDA+G on a $40 \times 40$ uniform 
grid in the parameter space spanned by $U$ and $J/U$. The spin state is determined by calculating $\Braket{G|S^2|G}$,
where $S^2$ is the total spin operator. The semiconductor gap is determined by the energy difference between
the bottom of conduction band and the top of valence band. The system is metallic when such semiconductor gap
becomes negative. Two typical  Coulomb interaction strength $U$ has been chosen to plot their
magnetization as a function of $J/U$, which is shown in Fig.~\ref{fig:sz},
and four typical points (red circle) have been chosen to plot their quasi-particle band structures and the 
probability of the atomic configurations, as shown in Fig.~\ref{fig:band}.
There are mainly three regions in the phase diagram: low spin ferromagnetic metal (LS-FM-M), low spin ferromagnetic insulator (LS-FM-I), 
and high spin ferromagnetic metal (HS-FM-M). The Coulomb interaction is much larger than the band width of
the flat band, as a consequence, the FM order can be easily formed and stabilized according to the Stoner's criteria.
As shown in Fig.~\ref{fig:sz}, with the increment of Hund's rule coupling
the FM polarization becomes stronger, and suddenly jumps at the phase boundary between LS phases and
HS phases indicating the corresponding transition is first order. 
In comparison,  we haven't find any stable AFM order in our calculations.

From the above discussion, we can draw a conclusion that the FM insulator phase can appear 
only when the following two conditions are satisfied: i) the effective SOC is big enough to split the band 
touching point at the $\Gamma$ point, ii) the FM exchange coupling is big enough to make the flat band 
around the chemical potential fully polarized. As discussed previously, the interaction parameters for 
LaCoO$_3$ are around $U=7.0$ eV and $J=0.6$ eV, with which our LDA+G calculation obtain quite robust 
FM insulator phase with Chern number 1. The corresponding quasi-particle band structure has been plotted 
in Fig.~\ref{fig:band}(c), where the semiconductor gap around $0.22$ eV lying between two bands with 
majority spin. The appearance of semiconductor behavior is mainly due to the effective SOC, which is 
greatly enhanced by the strong correlation effects in LaCoO$_3$. In this system, the effective SOC in 
the $e_{g}$ bands can be modified by the local correlation effects through two ways. The first effect is 
due to Hund's rule coupling $J$, which competes against the crystal field splitting between the $e_g$ and 
$t_{2g}$ bands and reduces the energy cost for virtual particle-hole excitations between them. Since the 
effective SOC in the $e_{g}$ bands is a second order effect caused by such virtual excitations, 
the Hund's rule coupling can then enhance the effective SOC.
The second effect is mainly due to the Coulomb interaction between different orbitals ($U^\prime =U-2J$),
which also enhances the effective SOC~\cite{liu:2008,du:2013} through the Hartree-Fock process especially when
the SOC splitting is between one almost fully occupied and one almost empty levels.
Although the Coulomb interaction is treated on the level of Gutzwiller approximation, we believe that 
the main physics of the interaction enhanced SOC has been well captured. 
We note that the Gutzwiller approximation only becomes exact in infinite dimension, where only the local 
correlation needs to be considered. The non-local correlation in 2D will be expected to reduce both
the semiconducting gap and the effective SOC. 
In the present study, the possible lattice distortion, which may also be enhanced by strong Coulomb 
interaction, has been neglected under the assumption that  the pinning force from the substrate is strong 
enough to prevent it from happening. While in more realistic systems, the above assumption
may not be well satisfied and further studies including the possible lattice distortion are needed,
which will be discussed elsewhere. 

As shown in Fig.~\ref{fig:phase}, in most of the phase region, the favorable spin state of bilayer LaCoO$_3$  
system is LS state. This is reasonable because the electron transfer induced by the charge mismatch 
increases the formal charge on Co $3d$ orbitals to be 
around $6.5$, which further stabilizes the LS state comparing with the bulk material. Without Hund's 
rule coupling the minimum Coulomb interaction strength required to open the semiconductor gap is around 
$3.0$ eV, which is far below the actual parameter for LaCoO$_3$ (7.0 eV), indicating the robustness of 
the predicted QAH phase in this system. The increasing of Hund's rule coupling will first enhance the 
effective SOC by reducing the energy cost for the virtual particle-hole excitation between $e_g$ and $t_{2g}$ 
bands and favors the FM insulator phase. While when $J/U$ is bigger than $0.05$, further increasing
$J$ will dramatically reduce the effective SOC and favors the FM metal phase. This is due to the fact that the
increment of $J$ always comes together with the decrement of inter-orbital repulsion $U^\prime$ (equals $U-2J$), 
which has the dominate effect on effective SOC in this phase region. 
With the actual Hubbard interaction $U=7.0$ eV, the effective SOC is always very large when it is in the LS state,
the FM insulator phase can only be destroyed by increasing the Hund's rule coupling $J$ to induce a spin state 
transition from LS to HS states. As plotted in Fig.~\ref{fig:band}(d), when the HS state 
is stabilized by strong enough Hund's rule coupling, two of the $t_{2g}$ bands are lifted across the Fermi level 
generating a FM metal phase. Therefore, the most important conclusion we can reach from the phase diagram is that 
with the reasonable $U$ and $J$ strength, as long as the LS state can be stabilized, the FM insulator phase 
with nonzero Chern number is always robust.

With the effective quasi-particle Hamiltonian obtained by LDA+G, we have also calculated the Chern number 
by Kubo formula~\cite{nagaosa:2010} for the FM insulator phase, which equals $1$ as we expected. 
The edge states along both zigzag and armchair type of edges are also calculated and plotted
in Figs.~\ref{fig:edge}(a, b), showing the typical chiral nature of the edge states in QAH states.  

% figure: the edge states of LaCoO3
\begin{figure}
\includegraphics[width=0.48\textwidth]{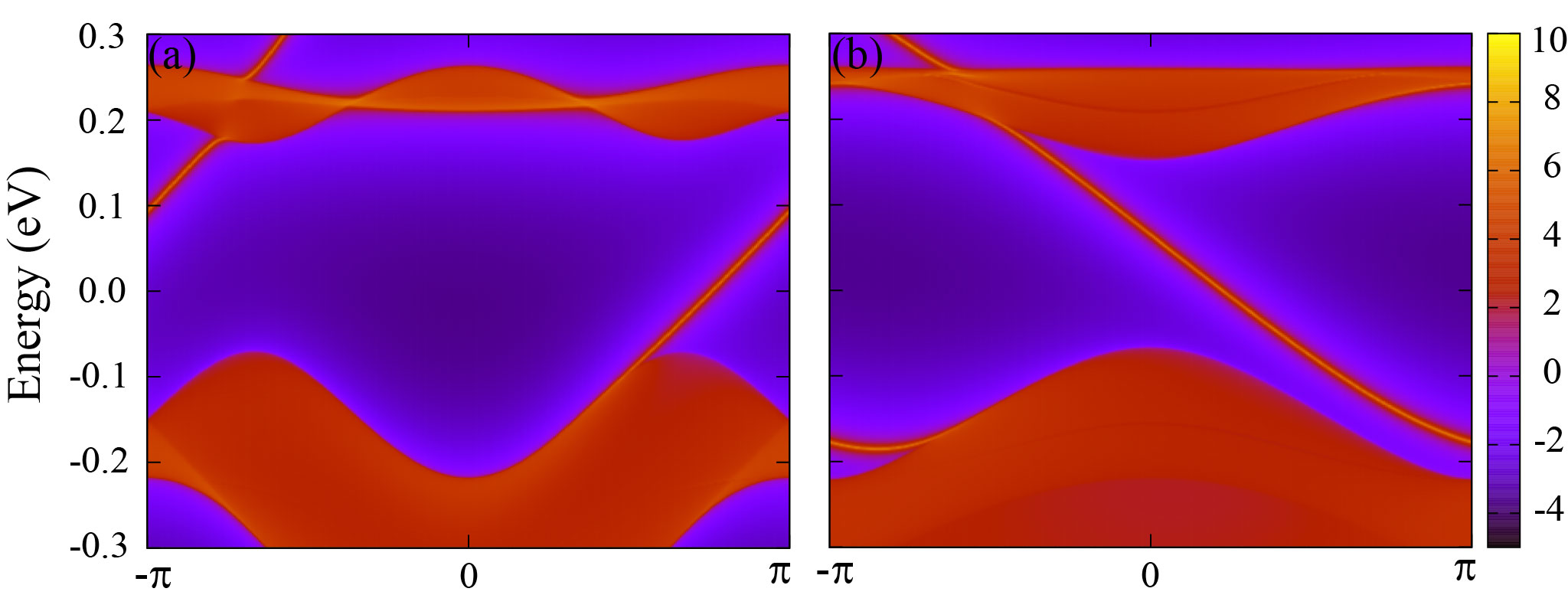}
\caption{(Color online). The local density of states for two identical edge modes of the buckled honeycomb lattice. 
         (a) is for zigzag edge and (b) is for armchair edge.}
\label{fig:edge}
\end{figure}

%% summary
\section{summary}\label{sec:sum}
In summary, we have studied the strongly correlated (111) bilayer heterostructure of LaCoO$_3$ by the LDA+G method. 
Our results verify that Coulomb interactions can largely enhance the effective SOC and stabilize a very robust 
LS-FM-I state due to the strong stoner instability in the topologically non-trivial flat band. 
The calculated Chern number $C=1$ and the edge states indicate possible QAH effect in this system. 
Besides, the strong correlation between the spin state of Co and the low energy band structure provides
another way to tune the topological properties. We believe that the fast development of oxide MBE technique provides
a great opportunity in this system to realize QAH effect.

%% acknowledge
\section{acknowledgments}\label{sec:ack}
We acknowledge the helpful discussions with professor Y. Ran and N. Nagaosa.
This work was supported by the NSF of China and by the 973 program of China (No. 2011CBA00108 and No.
2013CB921700). The calculations were preformed on TianHe-1(A), the National Supercomputer Center in Tianjin, China.

%% references
\bibliography{lacoo3}

\begin{thebibliography}{55}%
\makeatletter
\providecommand \@ifxundefined [1]{%
 \@ifx{#1\undefined}
}%
\providecommand \@ifnum [1]{%
 \ifnum #1\expandafter \@firstoftwo
 \else \expandafter \@secondoftwo
 \fi
}%
\providecommand \@ifx [1]{%
 \ifx #1\expandafter \@firstoftwo
 \else \expandafter \@secondoftwo
 \fi
}%
\providecommand \natexlab [1]{#1}%
\providecommand \enquote  [1]{``#1''}%
\providecommand \bibnamefont  [1]{#1}%
\providecommand \bibfnamefont [1]{#1}%
\providecommand \citenamefont [1]{#1}%
\providecommand \href@noop [0]{\@secondoftwo}%
\providecommand \href [0]{\begingroup \@sanitize@url \@href}%
\providecommand \@href[1]{\@@startlink{#1}\@@href}%
\providecommand \@@href[1]{\endgroup#1\@@endlink}%
\providecommand \@sanitize@url [0]{\catcode `\\12\catcode `\$12\catcode
  `\&12\catcode `\#12\catcode `\^12\catcode `\_12\catcode `\%12\relax}%
\providecommand \@@startlink[1]{}%
\providecommand \@@endlink[0]{}%
\providecommand \url  [0]{\begingroup\@sanitize@url \@url }%
\providecommand \@url [1]{\endgroup\@href {#1}{\urlprefix }}%
\providecommand \urlprefix  [0]{URL }%
\providecommand \Eprint [0]{\href }%
\providecommand \doibase [0]{http://dx.doi.org/}%
\providecommand \selectlanguage [0]{\@gobble}%
\providecommand \bibinfo  [0]{\@secondoftwo}%
\providecommand \bibfield  [0]{\@secondoftwo}%
\providecommand \translation [1]{[#1]}%
\providecommand \BibitemOpen [0]{}%
\providecommand \bibitemStop [0]{}%
\providecommand \bibitemNoStop [0]{.\EOS\space}%
\providecommand \EOS [0]{\spacefactor3000\relax}%
\providecommand \BibitemShut  [1]{\csname bibitem#1\endcsname}%
\let\auto@bib@innerbib\@empty
%</preamble>
\bibitem [{\citenamefont {Hasan}\ and\ \citenamefont
  {Kane}(2010)}]{hasan:2010}%
  \BibitemOpen
  \bibfield  {author} {\bibinfo {author} {\bibfnamefont {M.~Z.}\ \bibnamefont
  {Hasan}}\ and\ \bibinfo {author} {\bibfnamefont {C.~L.}\ \bibnamefont
  {Kane}},\ }\href {\doibase 10.1103/RevModPhys.82.3045} {\bibfield  {journal}
  {\bibinfo  {journal} {Rev. Mod. Phys.}\ }\textbf {\bibinfo {volume} {82}},\
  \bibinfo {pages} {3045} (\bibinfo {year} {2010})}\BibitemShut {NoStop}%
\bibitem [{\citenamefont {Qi}\ and\ \citenamefont {Zhang}(2011)}]{qi:2011}%
  \BibitemOpen
  \bibfield  {author} {\bibinfo {author} {\bibfnamefont {X.-L.}\ \bibnamefont
  {Qi}}\ and\ \bibinfo {author} {\bibfnamefont {S.-C.}\ \bibnamefont {Zhang}},\
  }\href {\doibase 10.1103/RevModPhys.83.1057} {\bibfield  {journal} {\bibinfo
  {journal} {Rev. Mod. Phys.}\ }\textbf {\bibinfo {volume} {83}},\ \bibinfo
  {pages} {1057} (\bibinfo {year} {2011})}\BibitemShut {NoStop}%
\bibitem [{\citenamefont {Xiao}\ \emph {et~al.}(2011)\citenamefont {Xiao},
  \citenamefont {Zhu}, \citenamefont {Ran}, \citenamefont {Nagaosa},\ and\
  \citenamefont {Okamoto}}]{xiao:2011}%
  \BibitemOpen
  \bibfield  {author} {\bibinfo {author} {\bibfnamefont {D.}~\bibnamefont
  {Xiao}}, \bibinfo {author} {\bibfnamefont {W.}~\bibnamefont {Zhu}}, \bibinfo
  {author} {\bibfnamefont {Y.}~\bibnamefont {Ran}}, \bibinfo {author}
  {\bibfnamefont {N.}~\bibnamefont {Nagaosa}}, \ and\ \bibinfo {author}
  {\bibfnamefont {S.}~\bibnamefont {Okamoto}},\ }\href
  {http://dx.doi.org/10.1038/ncomms1602} {\bibfield  {journal} {\bibinfo
  {journal} {Nat Cummun}\ }\textbf {\bibinfo {volume} {2}},\ \bibinfo {pages}
  {596} (\bibinfo {year} {2011})}\BibitemShut {NoStop}%
\bibitem [{\citenamefont {Bernevig}\ \emph {et~al.}(2006)\citenamefont
  {Bernevig}, \citenamefont {Hughes},\ and\ \citenamefont
  {Zhang}}]{bernevig:2006}%
  \BibitemOpen
  \bibfield  {author} {\bibinfo {author} {\bibfnamefont {B.~A.}\ \bibnamefont
  {Bernevig}}, \bibinfo {author} {\bibfnamefont {T.~L.}\ \bibnamefont
  {Hughes}}, \ and\ \bibinfo {author} {\bibfnamefont {S.-C.}\ \bibnamefont
  {Zhang}},\ }\href {\doibase 10.1126/science.1133734} {\bibfield  {journal}
  {\bibinfo  {journal} {Science}\ }\textbf {\bibinfo {volume} {314}},\ \bibinfo
  {pages} {1757} (\bibinfo {year} {2006})}\BibitemShut {NoStop}%
\bibitem [{\citenamefont {K\"{o}nig}\ \emph {et~al.}(2007)\citenamefont
  {K\"{o}nig}, \citenamefont {Wiedmann}, \citenamefont {Br\"{u}ne},
  \citenamefont {Roth}, \citenamefont {Buhmann}, \citenamefont {Molenkamp},
  \citenamefont {Qi},\ and\ \citenamefont {Zhang}}]{konig:2007}%
  \BibitemOpen
  \bibfield  {author} {\bibinfo {author} {\bibfnamefont {M.}~\bibnamefont
  {K\"{o}nig}}, \bibinfo {author} {\bibfnamefont {S.}~\bibnamefont {Wiedmann}},
  \bibinfo {author} {\bibfnamefont {C.}~\bibnamefont {Br\"{u}ne}}, \bibinfo
  {author} {\bibfnamefont {A.}~\bibnamefont {Roth}}, \bibinfo {author}
  {\bibfnamefont {H.}~\bibnamefont {Buhmann}}, \bibinfo {author} {\bibfnamefont
  {L.~W.}\ \bibnamefont {Molenkamp}}, \bibinfo {author} {\bibfnamefont {X.-L.}\
  \bibnamefont {Qi}}, \ and\ \bibinfo {author} {\bibfnamefont {S.-C.}\
  \bibnamefont {Zhang}},\ }\href {\doibase 10.1126/science.1148047} {\bibfield
  {journal} {\bibinfo  {journal} {Science}\ }\textbf {\bibinfo {volume}
  {318}},\ \bibinfo {pages} {766} (\bibinfo {year} {2007})}\BibitemShut
  {NoStop}%
\bibitem [{\citenamefont {Dai}\ \emph {et~al.}(2008)\citenamefont {Dai},
  \citenamefont {Hughes}, \citenamefont {Qi}, \citenamefont {Fang},\ and\
  \citenamefont {Zhang}}]{dai:2008}%
  \BibitemOpen
  \bibfield  {author} {\bibinfo {author} {\bibfnamefont {X.}~\bibnamefont
  {Dai}}, \bibinfo {author} {\bibfnamefont {T.~L.}\ \bibnamefont {Hughes}},
  \bibinfo {author} {\bibfnamefont {X.-L.}\ \bibnamefont {Qi}}, \bibinfo
  {author} {\bibfnamefont {Z.}~\bibnamefont {Fang}}, \ and\ \bibinfo {author}
  {\bibfnamefont {S.-C.}\ \bibnamefont {Zhang}},\ }\href {\doibase
  10.1103/PhysRevB.77.125319} {\bibfield  {journal} {\bibinfo  {journal} {Phys.
  Rev. B}\ }\textbf {\bibinfo {volume} {77}},\ \bibinfo {pages} {125319}
  (\bibinfo {year} {2008})}\BibitemShut {NoStop}%
\bibitem [{\citenamefont {Zhang}\ \emph {et~al.}(2009)\citenamefont {Zhang},
  \citenamefont {Liu}, \citenamefont {Qi}, \citenamefont {Dai}, \citenamefont
  {Fang},\ and\ \citenamefont {Zhang}}]{zhang:2009}%
  \BibitemOpen
  \bibfield  {author} {\bibinfo {author} {\bibfnamefont {H.}~\bibnamefont
  {Zhang}}, \bibinfo {author} {\bibfnamefont {C.-X.}\ \bibnamefont {Liu}},
  \bibinfo {author} {\bibfnamefont {X.-L.}\ \bibnamefont {Qi}}, \bibinfo
  {author} {\bibfnamefont {X.}~\bibnamefont {Dai}}, \bibinfo {author}
  {\bibfnamefont {Z.}~\bibnamefont {Fang}}, \ and\ \bibinfo {author}
  {\bibfnamefont {S.-C.}\ \bibnamefont {Zhang}},\ }\href
  {http://dx.doi.org/10.1038/nphys1270} {\bibfield  {journal} {\bibinfo
  {journal} {Nat Phys}\ }\textbf {\bibinfo {volume} {5}},\ \bibinfo {pages}
  {438} (\bibinfo {year} {2009})}\BibitemShut {NoStop}%
\bibitem [{\citenamefont {Xia}\ \emph {et~al.}(2009)\citenamefont {Xia},
  \citenamefont {Qian}, \citenamefont {Hsieh}, \citenamefont {Wray},
  \citenamefont {Pal}, \citenamefont {Lin}, \citenamefont {Bansil},
  \citenamefont {Grauer}, \citenamefont {Hor}, \citenamefont {Cava},\ and\
  \citenamefont {Hasan}}]{xia:2009}%
  \BibitemOpen
  \bibfield  {author} {\bibinfo {author} {\bibfnamefont {Y.}~\bibnamefont
  {Xia}}, \bibinfo {author} {\bibfnamefont {D.}~\bibnamefont {Qian}}, \bibinfo
  {author} {\bibfnamefont {D.}~\bibnamefont {Hsieh}}, \bibinfo {author}
  {\bibfnamefont {L.}~\bibnamefont {Wray}}, \bibinfo {author} {\bibfnamefont
  {A.}~\bibnamefont {Pal}}, \bibinfo {author} {\bibfnamefont {H.}~\bibnamefont
  {Lin}}, \bibinfo {author} {\bibfnamefont {A.}~\bibnamefont {Bansil}},
  \bibinfo {author} {\bibfnamefont {D.}~\bibnamefont {Grauer}}, \bibinfo
  {author} {\bibfnamefont {Y.~S.}\ \bibnamefont {Hor}}, \bibinfo {author}
  {\bibfnamefont {R.~J.}\ \bibnamefont {Cava}}, \ and\ \bibinfo {author}
  {\bibfnamefont {M.~Z.}\ \bibnamefont {Hasan}},\ }\href
  {http://dx.doi.org/10.1038/nphys1274} {\bibfield  {journal} {\bibinfo
  {journal} {Nat Phys}\ }\textbf {\bibinfo {volume} {5}},\ \bibinfo {pages}
  {398} (\bibinfo {year} {2009})}\BibitemShut {NoStop}%
\bibitem [{\citenamefont {Chen}\ \emph {et~al.}(2009)\citenamefont {Chen},
  \citenamefont {Analytis}, \citenamefont {Chu}, \citenamefont {Liu},
  \citenamefont {Mo}, \citenamefont {Qi}, \citenamefont {Zhang}, \citenamefont
  {Lu}, \citenamefont {Dai}, \citenamefont {Fang}, \citenamefont {Zhang},
  \citenamefont {Fisher}, \citenamefont {Hussain},\ and\ \citenamefont
  {Shen}}]{chen:2009}%
  \BibitemOpen
  \bibfield  {author} {\bibinfo {author} {\bibfnamefont {Y.~L.}\ \bibnamefont
  {Chen}}, \bibinfo {author} {\bibfnamefont {J.~G.}\ \bibnamefont {Analytis}},
  \bibinfo {author} {\bibfnamefont {J.-H.}\ \bibnamefont {Chu}}, \bibinfo
  {author} {\bibfnamefont {Z.~K.}\ \bibnamefont {Liu}}, \bibinfo {author}
  {\bibfnamefont {S.-K.}\ \bibnamefont {Mo}}, \bibinfo {author} {\bibfnamefont
  {X.~L.}\ \bibnamefont {Qi}}, \bibinfo {author} {\bibfnamefont {H.~J.}\
  \bibnamefont {Zhang}}, \bibinfo {author} {\bibfnamefont {D.~H.}\ \bibnamefont
  {Lu}}, \bibinfo {author} {\bibfnamefont {X.}~\bibnamefont {Dai}}, \bibinfo
  {author} {\bibfnamefont {Z.}~\bibnamefont {Fang}}, \bibinfo {author}
  {\bibfnamefont {S.~C.}\ \bibnamefont {Zhang}}, \bibinfo {author}
  {\bibfnamefont {I.~R.}\ \bibnamefont {Fisher}}, \bibinfo {author}
  {\bibfnamefont {Z.}~\bibnamefont {Hussain}}, \ and\ \bibinfo {author}
  {\bibfnamefont {Z.-X.}\ \bibnamefont {Shen}},\ }\href {\doibase
  10.1126/science.1173034} {\bibfield  {journal} {\bibinfo  {journal}
  {Science}\ }\textbf {\bibinfo {volume} {325}},\ \bibinfo {pages} {178}
  (\bibinfo {year} {2009})}\BibitemShut {NoStop}%
\bibitem [{\citenamefont {Izumi}\ \emph {et~al.}(2001)\citenamefont {Izumi},
  \citenamefont {Ogimoto}, \citenamefont {Konishi}, \citenamefont {Manako},
  \citenamefont {Kawasaki},\ and\ \citenamefont {Tokura}}]{izumi:2001}%
  \BibitemOpen
  \bibfield  {author} {\bibinfo {author} {\bibfnamefont {M.}~\bibnamefont
  {Izumi}}, \bibinfo {author} {\bibfnamefont {Y.}~\bibnamefont {Ogimoto}},
  \bibinfo {author} {\bibfnamefont {Y.}~\bibnamefont {Konishi}}, \bibinfo
  {author} {\bibfnamefont {T.}~\bibnamefont {Manako}}, \bibinfo {author}
  {\bibfnamefont {M.}~\bibnamefont {Kawasaki}}, \ and\ \bibinfo {author}
  {\bibfnamefont {Y.}~\bibnamefont {Tokura}},\ }\href {\doibase
  http://dx.doi.org/10.1016/S0921-5107(01)00569-4} {\bibfield  {journal}
  {\bibinfo  {journal} {Materials Science and Engineering: B}\ }\textbf
  {\bibinfo {volume} {84}},\ \bibinfo {pages} {53 } (\bibinfo {year}
  {2001})}\BibitemShut {NoStop}%
\bibitem [{\citenamefont {Ohtomo}\ \emph {et~al.}(2002)\citenamefont {Ohtomo},
  \citenamefont {Muller}, \citenamefont {Grazul},\ and\ \citenamefont
  {Hwang}}]{ohtomo:2002}%
  \BibitemOpen
  \bibfield  {author} {\bibinfo {author} {\bibfnamefont {A.}~\bibnamefont
  {Ohtomo}}, \bibinfo {author} {\bibfnamefont {D.~A.}\ \bibnamefont {Muller}},
  \bibinfo {author} {\bibfnamefont {J.~L.}\ \bibnamefont {Grazul}}, \ and\
  \bibinfo {author} {\bibfnamefont {H.~Y.}\ \bibnamefont {Hwang}},\ }\href
  {http://dx.doi.org/10.1038/nature00977} {\bibfield  {journal} {\bibinfo
  {journal} {Nature}\ }\textbf {\bibinfo {volume} {419}},\ \bibinfo {pages}
  {378} (\bibinfo {year} {2002})}\BibitemShut {NoStop}%
\bibitem [{\citenamefont {Ohtomo}\ and\ \citenamefont
  {Hwang}(2004)}]{ohtomo:2004}%
  \BibitemOpen
  \bibfield  {author} {\bibinfo {author} {\bibfnamefont {A.}~\bibnamefont
  {Ohtomo}}\ and\ \bibinfo {author} {\bibfnamefont {H.~Y.}\ \bibnamefont
  {Hwang}},\ }\href {http://dx.doi.org/10.1038/nature02308} {\bibfield
  {journal} {\bibinfo  {journal} {Nature}\ }\textbf {\bibinfo {volume} {427}},\
  \bibinfo {pages} {423} (\bibinfo {year} {2004})}\BibitemShut {NoStop}%
\bibitem [{\citenamefont {Castro~Neto}\ \emph {et~al.}(2009)\citenamefont
  {Castro~Neto}, \citenamefont {Guinea}, \citenamefont {Peres}, \citenamefont
  {Novoselov},\ and\ \citenamefont {Geim}}]{castro:2009}%
  \BibitemOpen
  \bibfield  {author} {\bibinfo {author} {\bibfnamefont {A.~H.}\ \bibnamefont
  {Castro~Neto}}, \bibinfo {author} {\bibfnamefont {F.}~\bibnamefont {Guinea}},
  \bibinfo {author} {\bibfnamefont {N.~M.~R.}\ \bibnamefont {Peres}}, \bibinfo
  {author} {\bibfnamefont {K.~S.}\ \bibnamefont {Novoselov}}, \ and\ \bibinfo
  {author} {\bibfnamefont {A.~K.}\ \bibnamefont {Geim}},\ }\href {\doibase
  10.1103/RevModPhys.81.109} {\bibfield  {journal} {\bibinfo  {journal} {Rev.
  Mod. Phys.}\ }\textbf {\bibinfo {volume} {81}},\ \bibinfo {pages} {109}
  (\bibinfo {year} {2009})}\BibitemShut {NoStop}%
\bibitem [{\citenamefont {Yang}\ \emph {et~al.}(2011)\citenamefont {Yang},
  \citenamefont {Zhu}, \citenamefont {Xiao}, \citenamefont {Okamoto},
  \citenamefont {Wang},\ and\ \citenamefont {Ran}}]{yang:2011}%
  \BibitemOpen
  \bibfield  {author} {\bibinfo {author} {\bibfnamefont {K.-Y.}\ \bibnamefont
  {Yang}}, \bibinfo {author} {\bibfnamefont {W.}~\bibnamefont {Zhu}}, \bibinfo
  {author} {\bibfnamefont {D.}~\bibnamefont {Xiao}}, \bibinfo {author}
  {\bibfnamefont {S.}~\bibnamefont {Okamoto}}, \bibinfo {author} {\bibfnamefont
  {Z.}~\bibnamefont {Wang}}, \ and\ \bibinfo {author} {\bibfnamefont
  {Y.}~\bibnamefont {Ran}},\ }\href {\doibase 10.1103/PhysRevB.84.201104}
  {\bibfield  {journal} {\bibinfo  {journal} {Phys. Rev. B}\ }\textbf {\bibinfo
  {volume} {84}},\ \bibinfo {pages} {201104} (\bibinfo {year}
  {2011})}\BibitemShut {NoStop}%
\bibitem [{\citenamefont {R\"uegg}\ and\ \citenamefont
  {Fiete}(2011)}]{andreas:2011}%
  \BibitemOpen
  \bibfield  {author} {\bibinfo {author} {\bibfnamefont {A.}~\bibnamefont
  {R\"uegg}}\ and\ \bibinfo {author} {\bibfnamefont {G.~A.}\ \bibnamefont
  {Fiete}},\ }\href {\doibase 10.1103/PhysRevB.84.201103} {\bibfield  {journal}
  {\bibinfo  {journal} {Phys. Rev. B}\ }\textbf {\bibinfo {volume} {84}},\
  \bibinfo {pages} {201103} (\bibinfo {year} {2011})}\BibitemShut {NoStop}%
\bibitem [{\citenamefont {R\"uegg}\ \emph {et~al.}(2012)\citenamefont
  {R\"uegg}, \citenamefont {Mitra}, \citenamefont {Demkov},\ and\ \citenamefont
  {Fiete}}]{andreas:2012}%
  \BibitemOpen
  \bibfield  {author} {\bibinfo {author} {\bibfnamefont {A.}~\bibnamefont
  {R\"uegg}}, \bibinfo {author} {\bibfnamefont {C.}~\bibnamefont {Mitra}},
  \bibinfo {author} {\bibfnamefont {A.~A.}\ \bibnamefont {Demkov}}, \ and\
  \bibinfo {author} {\bibfnamefont {G.~A.}\ \bibnamefont {Fiete}},\ }\href
  {\doibase 10.1103/PhysRevB.85.245131} {\bibfield  {journal} {\bibinfo
  {journal} {Phys. Rev. B}\ }\textbf {\bibinfo {volume} {85}},\ \bibinfo
  {pages} {245131} (\bibinfo {year} {2012})}\BibitemShut {NoStop}%
\bibitem [{\citenamefont {R\"uegg}\ \emph {et~al.}(2013)\citenamefont
  {R\"uegg}, \citenamefont {Mitra}, \citenamefont {Demkov},\ and\ \citenamefont
  {Fiete}}]{andreas:2013}%
  \BibitemOpen
  \bibfield  {author} {\bibinfo {author} {\bibfnamefont {A.}~\bibnamefont
  {R\"uegg}}, \bibinfo {author} {\bibfnamefont {C.}~\bibnamefont {Mitra}},
  \bibinfo {author} {\bibfnamefont {A.~A.}\ \bibnamefont {Demkov}}, \ and\
  \bibinfo {author} {\bibfnamefont {G.~A.}\ \bibnamefont {Fiete}},\ }\href
  {\doibase 10.1103/PhysRevB.88.115146} {\bibfield  {journal} {\bibinfo
  {journal} {Phys. Rev. B}\ }\textbf {\bibinfo {volume} {88}},\ \bibinfo
  {pages} {115146} (\bibinfo {year} {2013})}\BibitemShut {NoStop}%
\bibitem [{\citenamefont {Middey}\ \emph {et~al.}()\citenamefont {Middey},
  \citenamefont {Meyers}, \citenamefont {Doennig}, \citenamefont {Kareev},
  \citenamefont {Liu}, \citenamefont {Cao}, \citenamefont {Ryan}, \citenamefont
  {Pentcheva}, \citenamefont {Freeland},\ and\ \citenamefont
  {Chakhalian}}]{middey:2014}%
  \BibitemOpen
  \bibfield  {author} {\bibinfo {author} {\bibfnamefont {S.}~\bibnamefont
  {Middey}}, \bibinfo {author} {\bibfnamefont {D.}~\bibnamefont {Meyers}},
  \bibinfo {author} {\bibfnamefont {D.}~\bibnamefont {Doennig}}, \bibinfo
  {author} {\bibfnamefont {M.}~\bibnamefont {Kareev}}, \bibinfo {author}
  {\bibfnamefont {X.}~\bibnamefont {Liu}}, \bibinfo {author} {\bibfnamefont
  {Y.}~\bibnamefont {Cao}}, \bibinfo {author} {\bibfnamefont {P.~J.}\
  \bibnamefont {Ryan}}, \bibinfo {author} {\bibfnamefont {R.}~\bibnamefont
  {Pentcheva}}, \bibinfo {author} {\bibfnamefont {J.~W.}\ \bibnamefont
  {Freeland}}, \ and\ \bibinfo {author} {\bibfnamefont {J.}~\bibnamefont
  {Chakhalian}},\ }\href@noop {} {\ }\Eprint {http://arxiv.org/abs/1407.1570}
  {arXiv:1407.1570 [cond-mat]} \BibitemShut {NoStop}%
\bibitem [{\citenamefont {B\"unemann}\ \emph {et~al.}(1998)\citenamefont
  {B\"unemann}, \citenamefont {Weber},\ and\ \citenamefont
  {Gebhard}}]{bunemann:1998}%
  \BibitemOpen
  \bibfield  {author} {\bibinfo {author} {\bibfnamefont {J.}~\bibnamefont
  {B\"unemann}}, \bibinfo {author} {\bibfnamefont {W.}~\bibnamefont {Weber}}, \
  and\ \bibinfo {author} {\bibfnamefont {F.}~\bibnamefont {Gebhard}},\ }\href
  {\doibase 10.1103/PhysRevB.57.6896} {\bibfield  {journal} {\bibinfo
  {journal} {Phys. Rev. B}\ }\textbf {\bibinfo {volume} {57}},\ \bibinfo
  {pages} {6896} (\bibinfo {year} {1998})}\BibitemShut {NoStop}%
\bibitem [{\citenamefont {Deng}\ \emph {et~al.}(2009)\citenamefont {Deng},
  \citenamefont {Wang}, \citenamefont {Dai},\ and\ \citenamefont
  {Fang}}]{deng:2009}%
  \BibitemOpen
  \bibfield  {author} {\bibinfo {author} {\bibfnamefont {X.}~\bibnamefont
  {Deng}}, \bibinfo {author} {\bibfnamefont {L.}~\bibnamefont {Wang}}, \bibinfo
  {author} {\bibfnamefont {X.}~\bibnamefont {Dai}}, \ and\ \bibinfo {author}
  {\bibfnamefont {Z.}~\bibnamefont {Fang}},\ }\href {\doibase
  10.1103/PhysRevB.79.075114} {\bibfield  {journal} {\bibinfo  {journal} {Phys.
  Rev. B}\ }\textbf {\bibinfo {volume} {79}},\ \bibinfo {pages} {075114}
  (\bibinfo {year} {2009})}\BibitemShut {NoStop}%
\bibitem [{\citenamefont {Lanat\`a}\ \emph {et~al.}(2012)\citenamefont
  {Lanat\`a}, \citenamefont {Strand}, \citenamefont {Dai},\ and\ \citenamefont
  {Hellsing}}]{nicola:2012}%
  \BibitemOpen
  \bibfield  {author} {\bibinfo {author} {\bibfnamefont {N.}~\bibnamefont
  {Lanat\`a}}, \bibinfo {author} {\bibfnamefont {H.~U.~R.}\ \bibnamefont
  {Strand}}, \bibinfo {author} {\bibfnamefont {X.}~\bibnamefont {Dai}}, \ and\
  \bibinfo {author} {\bibfnamefont {B.}~\bibnamefont {Hellsing}},\ }\href
  {\doibase 10.1103/PhysRevB.85.035133} {\bibfield  {journal} {\bibinfo
  {journal} {Phys. Rev. B}\ }\textbf {\bibinfo {volume} {85}},\ \bibinfo
  {pages} {035133} (\bibinfo {year} {2012})}\BibitemShut {NoStop}%
\bibitem [{\citenamefont {Lu}\ \emph {et~al.}(2013)\citenamefont {Lu},
  \citenamefont {Zhao}, \citenamefont {Weng}, \citenamefont {Fang},\ and\
  \citenamefont {Dai}}]{lu:2013}%
  \BibitemOpen
  \bibfield  {author} {\bibinfo {author} {\bibfnamefont {F.}~\bibnamefont
  {Lu}}, \bibinfo {author} {\bibfnamefont {J.}~\bibnamefont {Zhao}}, \bibinfo
  {author} {\bibfnamefont {H.}~\bibnamefont {Weng}}, \bibinfo {author}
  {\bibfnamefont {Z.}~\bibnamefont {Fang}}, \ and\ \bibinfo {author}
  {\bibfnamefont {X.}~\bibnamefont {Dai}},\ }\href {\doibase
  10.1103/PhysRevLett.110.096401} {\bibfield  {journal} {\bibinfo  {journal}
  {Phys. Rev. Lett.}\ }\textbf {\bibinfo {volume} {110}},\ \bibinfo {pages}
  {096401} (\bibinfo {year} {2013})}\BibitemShut {NoStop}%
\bibitem [{\citenamefont {Kresse}\ and\ \citenamefont
  {Furthm\"uller}(1996)}]{kresse:1996}%
  \BibitemOpen
  \bibfield  {author} {\bibinfo {author} {\bibfnamefont {G.}~\bibnamefont
  {Kresse}}\ and\ \bibinfo {author} {\bibfnamefont {J.}~\bibnamefont
  {Furthm\"uller}},\ }\href {\doibase 10.1103/PhysRevB.54.11169} {\bibfield
  {journal} {\bibinfo  {journal} {Phys. Rev. B}\ }\textbf {\bibinfo {volume}
  {54}},\ \bibinfo {pages} {11169} (\bibinfo {year} {1996})}\BibitemShut
  {NoStop}%
\bibitem [{\citenamefont {Bl\"ochl}(1994)}]{blochl:1994}%
  \BibitemOpen
  \bibfield  {author} {\bibinfo {author} {\bibfnamefont {P.~E.}\ \bibnamefont
  {Bl\"ochl}},\ }\href {\doibase 10.1103/PhysRevB.50.17953} {\bibfield
  {journal} {\bibinfo  {journal} {Phys. Rev. B}\ }\textbf {\bibinfo {volume}
  {50}},\ \bibinfo {pages} {17953} (\bibinfo {year} {1994})}\BibitemShut
  {NoStop}%
\bibitem [{\citenamefont {Kresse}\ and\ \citenamefont
  {Joubert}(1999)}]{kresse:1999}%
  \BibitemOpen
  \bibfield  {author} {\bibinfo {author} {\bibfnamefont {G.}~\bibnamefont
  {Kresse}}\ and\ \bibinfo {author} {\bibfnamefont {D.}~\bibnamefont
  {Joubert}},\ }\href {\doibase 10.1103/PhysRevB.59.1758} {\bibfield  {journal}
  {\bibinfo  {journal} {Phys. Rev. B}\ }\textbf {\bibinfo {volume} {59}},\
  \bibinfo {pages} {1758} (\bibinfo {year} {1999})}\BibitemShut {NoStop}%
\bibitem [{\citenamefont {Perdew}\ \emph {et~al.}(1996)\citenamefont {Perdew},
  \citenamefont {Burke},\ and\ \citenamefont {Ernzerhof}}]{perdew:1996}%
  \BibitemOpen
  \bibfield  {author} {\bibinfo {author} {\bibfnamefont {J.~P.}\ \bibnamefont
  {Perdew}}, \bibinfo {author} {\bibfnamefont {K.}~\bibnamefont {Burke}}, \
  and\ \bibinfo {author} {\bibfnamefont {M.}~\bibnamefont {Ernzerhof}},\ }\href
  {\doibase 10.1103/PhysRevLett.77.3865} {\bibfield  {journal} {\bibinfo
  {journal} {Phys. Rev. Lett.}\ }\textbf {\bibinfo {volume} {77}},\ \bibinfo
  {pages} {3865} (\bibinfo {year} {1996})}\BibitemShut {NoStop}%
\bibitem [{\citenamefont {Marzari}\ \emph {et~al.}(2012)\citenamefont
  {Marzari}, \citenamefont {Mostofi}, \citenamefont {Yates}, \citenamefont
  {Souza},\ and\ \citenamefont {Vanderbilt}}]{marzari:2012}%
  \BibitemOpen
  \bibfield  {author} {\bibinfo {author} {\bibfnamefont {N.}~\bibnamefont
  {Marzari}}, \bibinfo {author} {\bibfnamefont {A.~A.}\ \bibnamefont
  {Mostofi}}, \bibinfo {author} {\bibfnamefont {J.~R.}\ \bibnamefont {Yates}},
  \bibinfo {author} {\bibfnamefont {I.}~\bibnamefont {Souza}}, \ and\ \bibinfo
  {author} {\bibfnamefont {D.}~\bibnamefont {Vanderbilt}},\ }\href {\doibase
  10.1103/RevModPhys.84.1419} {\bibfield  {journal} {\bibinfo  {journal} {Rev.
  Mod. Phys.}\ }\textbf {\bibinfo {volume} {84}},\ \bibinfo {pages} {1419}
  (\bibinfo {year} {2012})}\BibitemShut {NoStop}%
\bibitem [{\citenamefont {Mostofi}\ \emph {et~al.}(2008)\citenamefont
  {Mostofi}, \citenamefont {Yates}, \citenamefont {Lee}, \citenamefont {Souza},
  \citenamefont {Vanderbilt},\ and\ \citenamefont {Marzari}}]{mostofi:2008}%
  \BibitemOpen
  \bibfield  {author} {\bibinfo {author} {\bibfnamefont {A.~A.}\ \bibnamefont
  {Mostofi}}, \bibinfo {author} {\bibfnamefont {J.~R.}\ \bibnamefont {Yates}},
  \bibinfo {author} {\bibfnamefont {Y.-S.}\ \bibnamefont {Lee}}, \bibinfo
  {author} {\bibfnamefont {I.}~\bibnamefont {Souza}}, \bibinfo {author}
  {\bibfnamefont {D.}~\bibnamefont {Vanderbilt}}, \ and\ \bibinfo {author}
  {\bibfnamefont {N.}~\bibnamefont {Marzari}},\ }\href {\doibase
  http://dx.doi.org/10.1016/j.cpc.2007.11.016} {\bibfield  {journal} {\bibinfo
  {journal} {Computer Physics Communications}\ }\textbf {\bibinfo {volume}
  {178}},\ \bibinfo {pages} {685 } (\bibinfo {year} {2008})}\BibitemShut
  {NoStop}%
\bibitem [{\citenamefont {Sugano}\ \emph {et~al.}(1970)\citenamefont {Sugano},
  \citenamefont {Tanabe},\ and\ \citenamefont {Kamimura}}]{sugano:1970}%
  \BibitemOpen
  \bibfield  {author} {\bibinfo {author} {\bibfnamefont {S.}~\bibnamefont
  {Sugano}}, \bibinfo {author} {\bibfnamefont {Y.}~\bibnamefont {Tanabe}}, \
  and\ \bibinfo {author} {\bibfnamefont {H.}~\bibnamefont {Kamimura}},\
  }\href@noop {} {\emph {\bibinfo {title} {Multiplets of transition-metal ions
  in crystals}}}\ (\bibinfo  {publisher} {New York: Academic},\ \bibinfo {year}
  {1970})\BibitemShut {NoStop}%
\bibitem [{\citenamefont {Georges}\ \emph {et~al.}(2013)\citenamefont
  {Georges}, \citenamefont {Medici},\ and\ \citenamefont
  {Mravlje}}]{georges:2013}%
  \BibitemOpen
  \bibfield  {author} {\bibinfo {author} {\bibfnamefont {A.}~\bibnamefont
  {Georges}}, \bibinfo {author} {\bibfnamefont {L.~d.}\ \bibnamefont {Medici}},
  \ and\ \bibinfo {author} {\bibfnamefont {J.}~\bibnamefont {Mravlje}},\ }\href
  {\doibase 10.1146/annurev-conmatphys-020911-125045} {\bibfield  {journal}
  {\bibinfo  {journal} {Annual Review of Condensed Matter Physics}\ }\textbf
  {\bibinfo {volume} {4}},\ \bibinfo {pages} {137} (\bibinfo {year}
  {2013})}\BibitemShut {NoStop}%
\bibitem [{\citenamefont {de~Groot}\ \emph {et~al.}(1990)\citenamefont
  {de~Groot}, \citenamefont {Fuggle}, \citenamefont {Thole},\ and\
  \citenamefont {Sawatzky}}]{groot:1990}%
  \BibitemOpen
  \bibfield  {author} {\bibinfo {author} {\bibfnamefont {F.}~\bibnamefont
  {de~Groot}}, \bibinfo {author} {\bibfnamefont {J.}~\bibnamefont {Fuggle}},
  \bibinfo {author} {\bibfnamefont {B.}~\bibnamefont {Thole}}, \ and\ \bibinfo
  {author} {\bibfnamefont {G.}~\bibnamefont {Sawatzky}},\ }\href {\doibase
  10.1103/PhysRevB.42.5459} {\bibfield  {journal} {\bibinfo  {journal} {Phys.
  Rev. B}\ }\textbf {\bibinfo {volume} {42}},\ \bibinfo {pages} {5459}
  (\bibinfo {year} {1990})}\BibitemShut {NoStop}%
\bibitem [{\citenamefont {Anisimov}\ and\ \citenamefont
  {Izyumov}(2010)}]{anisimov:2010}%
  \BibitemOpen
  \bibfield  {author} {\bibinfo {author} {\bibfnamefont {V.}~\bibnamefont
  {Anisimov}}\ and\ \bibinfo {author} {\bibfnamefont {Y.}~\bibnamefont
  {Izyumov}},\ }\href {\doibase 10.1007/978-3-642-04826-5} {\emph {\bibinfo
  {title} {Electronic Structure of Strongly Correlated Materials}}}\ (\bibinfo
  {publisher} {Springer Berlin Heidelberg},\ \bibinfo {year}
  {2010})\BibitemShut {NoStop}%
\bibitem [{\citenamefont {Abbate}\ \emph {et~al.}(1994)\citenamefont {Abbate},
  \citenamefont {Potze}, \citenamefont {Sawatzky},\ and\ \citenamefont
  {Fujimori}}]{abbate:1994}%
  \BibitemOpen
  \bibfield  {author} {\bibinfo {author} {\bibfnamefont {M.}~\bibnamefont
  {Abbate}}, \bibinfo {author} {\bibfnamefont {R.}~\bibnamefont {Potze}},
  \bibinfo {author} {\bibfnamefont {G.~A.}\ \bibnamefont {Sawatzky}}, \ and\
  \bibinfo {author} {\bibfnamefont {A.}~\bibnamefont {Fujimori}},\ }\href
  {\doibase 10.1103/PhysRevB.49.7210} {\bibfield  {journal} {\bibinfo
  {journal} {Phys. Rev. B}\ }\textbf {\bibinfo {volume} {49}},\ \bibinfo
  {pages} {7210} (\bibinfo {year} {1994})}\BibitemShut {NoStop}%
\bibitem [{\citenamefont {Korotin}\ \emph {et~al.}(1996)\citenamefont
  {Korotin}, \citenamefont {Ezhov}, \citenamefont {Solovyev}, \citenamefont
  {Anisimov}, \citenamefont {Khomskii},\ and\ \citenamefont
  {Sawatzky}}]{korotin:1996}%
  \BibitemOpen
  \bibfield  {author} {\bibinfo {author} {\bibfnamefont {M.~A.}\ \bibnamefont
  {Korotin}}, \bibinfo {author} {\bibfnamefont {S.~Y.}\ \bibnamefont {Ezhov}},
  \bibinfo {author} {\bibfnamefont {I.~V.}\ \bibnamefont {Solovyev}}, \bibinfo
  {author} {\bibfnamefont {V.~I.}\ \bibnamefont {Anisimov}}, \bibinfo {author}
  {\bibfnamefont {D.~I.}\ \bibnamefont {Khomskii}}, \ and\ \bibinfo {author}
  {\bibfnamefont {G.~A.}\ \bibnamefont {Sawatzky}},\ }\href {\doibase
  10.1103/PhysRevB.54.5309} {\bibfield  {journal} {\bibinfo  {journal} {Phys.
  Rev. B}\ }\textbf {\bibinfo {volume} {54}},\ \bibinfo {pages} {5309}
  (\bibinfo {year} {1996})}\BibitemShut {NoStop}%
\bibitem [{\citenamefont {Yamaguchi}\ \emph {et~al.}(1997)\citenamefont
  {Yamaguchi}, \citenamefont {Okimoto},\ and\ \citenamefont
  {Tokura}}]{yamaguchi:1997}%
  \BibitemOpen
  \bibfield  {author} {\bibinfo {author} {\bibfnamefont {S.}~\bibnamefont
  {Yamaguchi}}, \bibinfo {author} {\bibfnamefont {Y.}~\bibnamefont {Okimoto}},
  \ and\ \bibinfo {author} {\bibfnamefont {Y.}~\bibnamefont {Tokura}},\ }\href
  {\doibase 10.1103/PhysRevB.55.R8666} {\bibfield  {journal} {\bibinfo
  {journal} {Phys. Rev. B}\ }\textbf {\bibinfo {volume} {55}},\ \bibinfo
  {pages} {R8666} (\bibinfo {year} {1997})}\BibitemShut {NoStop}%
\bibitem [{\citenamefont {Asai}\ \emph {et~al.}(1998)\citenamefont {Asai},
  \citenamefont {Yoneda}, \citenamefont {Yokokura}, \citenamefont {Tranquada},
  \citenamefont {Shirane},\ and\ \citenamefont {Kohn}}]{asai:1998}%
  \BibitemOpen
  \bibfield  {author} {\bibinfo {author} {\bibfnamefont {K.}~\bibnamefont
  {Asai}}, \bibinfo {author} {\bibfnamefont {A.}~\bibnamefont {Yoneda}},
  \bibinfo {author} {\bibfnamefont {O.}~\bibnamefont {Yokokura}}, \bibinfo
  {author} {\bibfnamefont {J.}~\bibnamefont {Tranquada}}, \bibinfo {author}
  {\bibfnamefont {G.}~\bibnamefont {Shirane}}, \ and\ \bibinfo {author}
  {\bibfnamefont {K.}~\bibnamefont {Kohn}},\ }\href {\doibase
  10.1143/JPSJ.67.290} {\bibfield  {journal} {\bibinfo  {journal} {Journal of
  the Physical Society of Japan}\ }\textbf {\bibinfo {volume} {67}},\ \bibinfo
  {pages} {290} (\bibinfo {year} {1998})}\BibitemShut {NoStop}%
\bibitem [{\citenamefont {Yamaguchi}\ \emph {et~al.}(1996)\citenamefont
  {Yamaguchi}, \citenamefont {Okimoto}, \citenamefont {Taniguchi},\ and\
  \citenamefont {Tokura}}]{yamaguchi:1996}%
  \BibitemOpen
  \bibfield  {author} {\bibinfo {author} {\bibfnamefont {S.}~\bibnamefont
  {Yamaguchi}}, \bibinfo {author} {\bibfnamefont {Y.}~\bibnamefont {Okimoto}},
  \bibinfo {author} {\bibfnamefont {H.}~\bibnamefont {Taniguchi}}, \ and\
  \bibinfo {author} {\bibfnamefont {Y.}~\bibnamefont {Tokura}},\ }\href
  {\doibase 10.1103/PhysRevB.53.R2926} {\bibfield  {journal} {\bibinfo
  {journal} {Phys. Rev. B}\ }\textbf {\bibinfo {volume} {53}},\ \bibinfo
  {pages} {R2926} (\bibinfo {year} {1996})}\BibitemShut {NoStop}%
\bibitem [{\citenamefont {Tokura}\ \emph {et~al.}(1998)\citenamefont {Tokura},
  \citenamefont {Okimoto}, \citenamefont {Yamaguchi}, \citenamefont
  {Taniguchi}, \citenamefont {Kimura},\ and\ \citenamefont
  {Takagi}}]{tokura:1998}%
  \BibitemOpen
  \bibfield  {author} {\bibinfo {author} {\bibfnamefont {Y.}~\bibnamefont
  {Tokura}}, \bibinfo {author} {\bibfnamefont {Y.}~\bibnamefont {Okimoto}},
  \bibinfo {author} {\bibfnamefont {S.}~\bibnamefont {Yamaguchi}}, \bibinfo
  {author} {\bibfnamefont {H.}~\bibnamefont {Taniguchi}}, \bibinfo {author}
  {\bibfnamefont {T.}~\bibnamefont {Kimura}}, \ and\ \bibinfo {author}
  {\bibfnamefont {H.}~\bibnamefont {Takagi}},\ }\href {\doibase
  10.1103/PhysRevB.58.R1699} {\bibfield  {journal} {\bibinfo  {journal} {Phys.
  Rev. B}\ }\textbf {\bibinfo {volume} {58}},\ \bibinfo {pages} {R1699}
  (\bibinfo {year} {1998})}\BibitemShut {NoStop}%
\bibitem [{\citenamefont {Imada}\ \emph {et~al.}(1998)\citenamefont {Imada},
  \citenamefont {Fujimori},\ and\ \citenamefont {Tokura}}]{imada:1998}%
  \BibitemOpen
  \bibfield  {author} {\bibinfo {author} {\bibfnamefont {M.}~\bibnamefont
  {Imada}}, \bibinfo {author} {\bibfnamefont {A.}~\bibnamefont {Fujimori}}, \
  and\ \bibinfo {author} {\bibfnamefont {Y.}~\bibnamefont {Tokura}},\ }\href
  {\doibase 10.1103/RevModPhys.70.1039} {\bibfield  {journal} {\bibinfo
  {journal} {Rev. Mod. Phys.}\ }\textbf {\bibinfo {volume} {70}},\ \bibinfo
  {pages} {1039} (\bibinfo {year} {1998})}\BibitemShut {NoStop}%
\bibitem [{\citenamefont {Radaelli}\ and\ \citenamefont
  {Cheong}(2002)}]{radaelli:2002}%
  \BibitemOpen
  \bibfield  {author} {\bibinfo {author} {\bibfnamefont {P.~G.}\ \bibnamefont
  {Radaelli}}\ and\ \bibinfo {author} {\bibfnamefont {S.-W.}\ \bibnamefont
  {Cheong}},\ }\href {\doibase 10.1103/PhysRevB.66.094408} {\bibfield
  {journal} {\bibinfo  {journal} {Phys. Rev. B}\ }\textbf {\bibinfo {volume}
  {66}},\ \bibinfo {pages} {094408} (\bibinfo {year} {2002})}\BibitemShut
  {NoStop}%
\bibitem [{\citenamefont {Zobel}\ \emph {et~al.}(2002)\citenamefont {Zobel},
  \citenamefont {Kriener}, \citenamefont {Bruns}, \citenamefont {Baier},
  \citenamefont {Gr\"uninger}, \citenamefont {Lorenz}, \citenamefont
  {Reutler},\ and\ \citenamefont {Revcolevschi}}]{zobel:2002}%
  \BibitemOpen
  \bibfield  {author} {\bibinfo {author} {\bibfnamefont {C.}~\bibnamefont
  {Zobel}}, \bibinfo {author} {\bibfnamefont {M.}~\bibnamefont {Kriener}},
  \bibinfo {author} {\bibfnamefont {D.}~\bibnamefont {Bruns}}, \bibinfo
  {author} {\bibfnamefont {J.}~\bibnamefont {Baier}}, \bibinfo {author}
  {\bibfnamefont {M.}~\bibnamefont {Gr\"uninger}}, \bibinfo {author}
  {\bibfnamefont {T.}~\bibnamefont {Lorenz}}, \bibinfo {author} {\bibfnamefont
  {P.}~\bibnamefont {Reutler}}, \ and\ \bibinfo {author} {\bibfnamefont
  {A.}~\bibnamefont {Revcolevschi}},\ }\href {\doibase
  10.1103/PhysRevB.66.020402} {\bibfield  {journal} {\bibinfo  {journal} {Phys.
  Rev. B}\ }\textbf {\bibinfo {volume} {66}},\ \bibinfo {pages} {020402}
  (\bibinfo {year} {2002})}\BibitemShut {NoStop}%
\bibitem [{\citenamefont {Kn\'\i\ifmmode~\check{z}\else \v{z}\fi{}ek}\ \emph
  {et~al.}(2005)\citenamefont {Kn\'\i\ifmmode~\check{z}\else \v{z}\fi{}ek},
  \citenamefont {Nov\'ak},\ and\ \citenamefont {Jir\'ak}}]{knizek:2005}%
  \BibitemOpen
  \bibfield  {author} {\bibinfo {author} {\bibfnamefont {K.}~\bibnamefont
  {Kn\'\i\ifmmode~\check{z}\else \v{z}\fi{}ek}}, \bibinfo {author}
  {\bibfnamefont {P.}~\bibnamefont {Nov\'ak}}, \ and\ \bibinfo {author}
  {\bibfnamefont {Z.}~\bibnamefont {Jir\'ak}},\ }\href {\doibase
  10.1103/PhysRevB.71.054420} {\bibfield  {journal} {\bibinfo  {journal} {Phys.
  Rev. B}\ }\textbf {\bibinfo {volume} {71}},\ \bibinfo {pages} {054420}
  (\bibinfo {year} {2005})}\BibitemShut {NoStop}%
\bibitem [{\citenamefont {Haverkort}\ \emph {et~al.}(2006)\citenamefont
  {Haverkort}, \citenamefont {Hu}, \citenamefont {Cezar}, \citenamefont
  {Burnus}, \citenamefont {Hartmann}, \citenamefont {Reuther}, \citenamefont
  {Zobel}, \citenamefont {Lorenz}, \citenamefont {Tanaka}, \citenamefont
  {Brookes}, \citenamefont {Hsieh}, \citenamefont {Lin}, \citenamefont {Chen},\
  and\ \citenamefont {Tjeng}}]{haverkort:2006}%
  \BibitemOpen
  \bibfield  {author} {\bibinfo {author} {\bibfnamefont {M.~W.}\ \bibnamefont
  {Haverkort}}, \bibinfo {author} {\bibfnamefont {Z.}~\bibnamefont {Hu}},
  \bibinfo {author} {\bibfnamefont {J.~C.}\ \bibnamefont {Cezar}}, \bibinfo
  {author} {\bibfnamefont {T.}~\bibnamefont {Burnus}}, \bibinfo {author}
  {\bibfnamefont {H.}~\bibnamefont {Hartmann}}, \bibinfo {author}
  {\bibfnamefont {M.}~\bibnamefont {Reuther}}, \bibinfo {author} {\bibfnamefont
  {C.}~\bibnamefont {Zobel}}, \bibinfo {author} {\bibfnamefont
  {T.}~\bibnamefont {Lorenz}}, \bibinfo {author} {\bibfnamefont
  {A.}~\bibnamefont {Tanaka}}, \bibinfo {author} {\bibfnamefont {N.~B.}\
  \bibnamefont {Brookes}}, \bibinfo {author} {\bibfnamefont {H.~H.}\
  \bibnamefont {Hsieh}}, \bibinfo {author} {\bibfnamefont {H.-J.}\ \bibnamefont
  {Lin}}, \bibinfo {author} {\bibfnamefont {C.~T.}\ \bibnamefont {Chen}}, \
  and\ \bibinfo {author} {\bibfnamefont {L.~H.}\ \bibnamefont {Tjeng}},\ }\href
  {\doibase 10.1103/PhysRevLett.97.176405} {\bibfield  {journal} {\bibinfo
  {journal} {Phys. Rev. Lett.}\ }\textbf {\bibinfo {volume} {97}},\ \bibinfo
  {pages} {176405} (\bibinfo {year} {2006})}\BibitemShut {NoStop}%
\bibitem [{\citenamefont {Klie}\ \emph {et~al.}(2007)\citenamefont {Klie},
  \citenamefont {Zheng}, \citenamefont {Zhu}, \citenamefont {Varela},
  \citenamefont {Wu},\ and\ \citenamefont {Leighton}}]{klie:2007}%
  \BibitemOpen
  \bibfield  {author} {\bibinfo {author} {\bibfnamefont {R.~F.}\ \bibnamefont
  {Klie}}, \bibinfo {author} {\bibfnamefont {J.~C.}\ \bibnamefont {Zheng}},
  \bibinfo {author} {\bibfnamefont {Y.}~\bibnamefont {Zhu}}, \bibinfo {author}
  {\bibfnamefont {M.}~\bibnamefont {Varela}}, \bibinfo {author} {\bibfnamefont
  {J.}~\bibnamefont {Wu}}, \ and\ \bibinfo {author} {\bibfnamefont
  {C.}~\bibnamefont {Leighton}},\ }\href {\doibase
  10.1103/PhysRevLett.99.047203} {\bibfield  {journal} {\bibinfo  {journal}
  {Phys. Rev. Lett.}\ }\textbf {\bibinfo {volume} {99}},\ \bibinfo {pages}
  {047203} (\bibinfo {year} {2007})}\BibitemShut {NoStop}%
\bibitem [{\citenamefont {Pandey}\ \emph {et~al.}(2008)\citenamefont {Pandey},
  \citenamefont {Kumar}, \citenamefont {Patil}, \citenamefont {Medicherla},
  \citenamefont {Singh}, \citenamefont {Maiti}, \citenamefont {Prabhakaran},
  \citenamefont {Boothroyd},\ and\ \citenamefont {Pimpale}}]{pandey:2008}%
  \BibitemOpen
  \bibfield  {author} {\bibinfo {author} {\bibfnamefont {S.~K.}\ \bibnamefont
  {Pandey}}, \bibinfo {author} {\bibfnamefont {A.}~\bibnamefont {Kumar}},
  \bibinfo {author} {\bibfnamefont {S.}~\bibnamefont {Patil}}, \bibinfo
  {author} {\bibfnamefont {V.~R.~R.}\ \bibnamefont {Medicherla}}, \bibinfo
  {author} {\bibfnamefont {R.~S.}\ \bibnamefont {Singh}}, \bibinfo {author}
  {\bibfnamefont {K.}~\bibnamefont {Maiti}}, \bibinfo {author} {\bibfnamefont
  {D.}~\bibnamefont {Prabhakaran}}, \bibinfo {author} {\bibfnamefont {A.~T.}\
  \bibnamefont {Boothroyd}}, \ and\ \bibinfo {author} {\bibfnamefont {A.~V.}\
  \bibnamefont {Pimpale}},\ }\href {\doibase 10.1103/PhysRevB.77.045123}
  {\bibfield  {journal} {\bibinfo  {journal} {Phys. Rev. B}\ }\textbf {\bibinfo
  {volume} {77}},\ \bibinfo {pages} {045123} (\bibinfo {year}
  {2008})}\BibitemShut {NoStop}%
\bibitem [{\citenamefont {Craco}\ and\ \citenamefont
  {M\"uller-Hartmann}(2008)}]{craco:2008}%
  \BibitemOpen
  \bibfield  {author} {\bibinfo {author} {\bibfnamefont {L.}~\bibnamefont
  {Craco}}\ and\ \bibinfo {author} {\bibfnamefont {E.}~\bibnamefont
  {M\"uller-Hartmann}},\ }\href {\doibase 10.1103/PhysRevB.77.045130}
  {\bibfield  {journal} {\bibinfo  {journal} {Phys. Rev. B}\ }\textbf {\bibinfo
  {volume} {77}},\ \bibinfo {pages} {045130} (\bibinfo {year}
  {2008})}\BibitemShut {NoStop}%
\bibitem [{\citenamefont {Hsu}\ \emph {et~al.}(2009)\citenamefont {Hsu},
  \citenamefont {Umemoto}, \citenamefont {Cococcioni},\ and\ \citenamefont
  {Wentzcovitch}}]{hsu:2009}%
  \BibitemOpen
  \bibfield  {author} {\bibinfo {author} {\bibfnamefont {H.}~\bibnamefont
  {Hsu}}, \bibinfo {author} {\bibfnamefont {K.}~\bibnamefont {Umemoto}},
  \bibinfo {author} {\bibfnamefont {M.}~\bibnamefont {Cococcioni}}, \ and\
  \bibinfo {author} {\bibfnamefont {R.}~\bibnamefont {Wentzcovitch}},\ }\href
  {\doibase 10.1103/PhysRevB.79.125124} {\bibfield  {journal} {\bibinfo
  {journal} {Phys. Rev. B}\ }\textbf {\bibinfo {volume} {79}},\ \bibinfo
  {pages} {125124} (\bibinfo {year} {2009})}\BibitemShut {NoStop}%
\bibitem [{\citenamefont {Eder}(2010)}]{eder:2010}%
  \BibitemOpen
  \bibfield  {author} {\bibinfo {author} {\bibfnamefont {R.}~\bibnamefont
  {Eder}},\ }\href {\doibase 10.1103/PhysRevB.81.035101} {\bibfield  {journal}
  {\bibinfo  {journal} {Phys. Rev. B}\ }\textbf {\bibinfo {volume} {81}},\
  \bibinfo {pages} {035101} (\bibinfo {year} {2010})}\BibitemShut {NoStop}%
\bibitem [{\citenamefont {Kune\ifmmode~\check{s}\else \v{s}\fi{}}\ and\
  \citenamefont {K\ifmmode~\check{r}\else \v{r}\fi{}\'apek}(2011)}]{kune:2011}%
  \BibitemOpen
  \bibfield  {author} {\bibinfo {author} {\bibfnamefont {J.}~\bibnamefont
  {Kune\ifmmode~\check{s}\else \v{s}\fi{}}}\ and\ \bibinfo {author}
  {\bibfnamefont {V.}~\bibnamefont {K\ifmmode~\check{r}\else
  \v{r}\fi{}\'apek}},\ }\href {\doibase 10.1103/PhysRevLett.106.256401}
  {\bibfield  {journal} {\bibinfo  {journal} {Phys. Rev. Lett.}\ }\textbf
  {\bibinfo {volume} {106}},\ \bibinfo {pages} {256401} (\bibinfo {year}
  {2011})}\BibitemShut {NoStop}%
\bibitem [{\citenamefont {K\ifmmode~\check{r}\else \v{r}\fi{}\'apek}\ \emph
  {et~al.}(2012)\citenamefont {K\ifmmode~\check{r}\else \v{r}\fi{}\'apek},
  \citenamefont {Nov\'ak}, \citenamefont {Kune\ifmmode~\check{s}\else
  \v{s}\fi{}}, \citenamefont {Novoselov}, \citenamefont {Korotin},\ and\
  \citenamefont {Anisimov}}]{krapek:2012}%
  \BibitemOpen
  \bibfield  {author} {\bibinfo {author} {\bibfnamefont {V.}~\bibnamefont
  {K\ifmmode~\check{r}\else \v{r}\fi{}\'apek}}, \bibinfo {author}
  {\bibfnamefont {P.}~\bibnamefont {Nov\'ak}}, \bibinfo {author} {\bibfnamefont
  {J.}~\bibnamefont {Kune\ifmmode~\check{s}\else \v{s}\fi{}}}, \bibinfo
  {author} {\bibfnamefont {D.}~\bibnamefont {Novoselov}}, \bibinfo {author}
  {\bibfnamefont {D.~M.}\ \bibnamefont {Korotin}}, \ and\ \bibinfo {author}
  {\bibfnamefont {V.~I.}\ \bibnamefont {Anisimov}},\ }\href
  {http://link.aps.org/doi/10.1103/PhysRevB.86.195104} {\bibfield  {journal}
  {\bibinfo  {journal} {Phys. Rev. B}\ }\textbf {\bibinfo {volume} {86}},\
  \bibinfo {pages} {195104} (\bibinfo {year} {2012})}\BibitemShut {NoStop}%
\bibitem [{\citenamefont {Zhang}\ \emph {et~al.}(2012)\citenamefont {Zhang},
  \citenamefont {Gorelov}, \citenamefont {Koch},\ and\ \citenamefont
  {Pavarini}}]{zhang:2012}%
  \BibitemOpen
  \bibfield  {author} {\bibinfo {author} {\bibfnamefont {G.}~\bibnamefont
  {Zhang}}, \bibinfo {author} {\bibfnamefont {E.}~\bibnamefont {Gorelov}},
  \bibinfo {author} {\bibfnamefont {E.}~\bibnamefont {Koch}}, \ and\ \bibinfo
  {author} {\bibfnamefont {E.}~\bibnamefont {Pavarini}},\ }\href {\doibase
  10.1103/PhysRevB.86.184413} {\bibfield  {journal} {\bibinfo  {journal} {Phys.
  Rev. B}\ }\textbf {\bibinfo {volume} {86}},\ \bibinfo {pages} {184413}
  (\bibinfo {year} {2012})}\BibitemShut {NoStop}%
\bibitem [{\citenamefont {Sun}\ \emph {et~al.}(2009)\citenamefont {Sun},
  \citenamefont {Yao}, \citenamefont {Fradkin},\ and\ \citenamefont
  {Kivelson}}]{sunkai:2009}%
  \BibitemOpen
  \bibfield  {author} {\bibinfo {author} {\bibfnamefont {K.}~\bibnamefont
  {Sun}}, \bibinfo {author} {\bibfnamefont {H.}~\bibnamefont {Yao}}, \bibinfo
  {author} {\bibfnamefont {E.}~\bibnamefont {Fradkin}}, \ and\ \bibinfo
  {author} {\bibfnamefont {S.~A.}\ \bibnamefont {Kivelson}},\ }\href {\doibase
  10.1103/PhysRevLett.103.046811} {\bibfield  {journal} {\bibinfo  {journal}
  {Phys. Rev. Lett.}\ }\textbf {\bibinfo {volume} {103}},\ \bibinfo {pages}
  {046811} (\bibinfo {year} {2009})}\BibitemShut {NoStop}%
\bibitem [{\citenamefont {Liu}\ \emph {et~al.}(2008)\citenamefont {Liu},
  \citenamefont {Antonov}, \citenamefont {Jepsen},\ and\ \citenamefont
  {Andersen.}}]{liu:2008}%
  \BibitemOpen
  \bibfield  {author} {\bibinfo {author} {\bibfnamefont {G.-Q.}\ \bibnamefont
  {Liu}}, \bibinfo {author} {\bibfnamefont {V.~N.}\ \bibnamefont {Antonov}},
  \bibinfo {author} {\bibfnamefont {O.}~\bibnamefont {Jepsen}}, \ and\ \bibinfo
  {author} {\bibfnamefont {O.~K.}\ \bibnamefont {Andersen.}},\ }\href {\doibase
  10.1103/PhysRevLett.101.026408} {\bibfield  {journal} {\bibinfo  {journal}
  {Phys. Rev. Lett.}\ }\textbf {\bibinfo {volume} {101}},\ \bibinfo {pages}
  {026408} (\bibinfo {year} {2008})}\BibitemShut {NoStop}%
\bibitem [{\citenamefont {Du}\ \emph {et~al.}(2013)\citenamefont {Du},
  \citenamefont {Huang},\ and\ \citenamefont {Dai}}]{du:2013}%
  \BibitemOpen
  \bibfield  {author} {\bibinfo {author} {\bibfnamefont {L.}~\bibnamefont
  {Du}}, \bibinfo {author} {\bibfnamefont {L.}~\bibnamefont {Huang}}, \ and\
  \bibinfo {author} {\bibfnamefont {X.}~\bibnamefont {Dai}},\ }\href {\doibase
  10.1140/epjb/e2013-31024-6} {\bibfield  {journal} {\bibinfo  {journal} {The
  European Physical Journal B}\ }\textbf {\bibinfo {volume} {86}},\ \bibinfo
  {pages} {1} (\bibinfo {year} {2013})}\BibitemShut {NoStop}%
\bibitem [{\citenamefont {Nagaosa}\ \emph {et~al.}(2010)\citenamefont
  {Nagaosa}, \citenamefont {Sinova}, \citenamefont {Onoda}, \citenamefont
  {MacDonald},\ and\ \citenamefont {Ong}}]{nagaosa:2010}%
  \BibitemOpen
  \bibfield  {author} {\bibinfo {author} {\bibfnamefont {N.}~\bibnamefont
  {Nagaosa}}, \bibinfo {author} {\bibfnamefont {J.}~\bibnamefont {Sinova}},
  \bibinfo {author} {\bibfnamefont {S.}~\bibnamefont {Onoda}}, \bibinfo
  {author} {\bibfnamefont {A.~H.}\ \bibnamefont {MacDonald}}, \ and\ \bibinfo
  {author} {\bibfnamefont {N.~P.}\ \bibnamefont {Ong}},\ }\href {\doibase
  10.1103/RevModPhys.82.1539} {\bibfield  {journal} {\bibinfo  {journal} {Rev.
  Mod. Phys.}\ }\textbf {\bibinfo {volume} {82}},\ \bibinfo {pages} {1539}
  (\bibinfo {year} {2010})}\BibitemShut {NoStop}%
\end{thebibliography}%

\end{document}